\documentclass[twoside,twocolumn,9pt]{article}
\usepackage{extsizes}
\usepackage[super,sort&compress,comma]{natbib}
\usepackage[version=3]{mhchem}
\usepackage[left=1.5cm, right=1.5cm, top=1.785cm, bottom=2.0cm]{geometry}
\usepackage{balance}
\usepackage{times,mathptmx}
\usepackage{sectsty}
\usepackage{graphicx}
\usepackage{lastpage}
\usepackage[format=plain,justification=justified,singlelinecheck=false,font={stretch=1.125,small,sf},labelfont=bf,labelsep=space]{caption}
\usepackage{float}
\usepackage{fancyhdr}
\usepackage{fnpos}
\usepackage{array}
\usepackage{droidsans}
\usepackage[para]{threeparttable}
\usepackage{charter}
\usepackage[T1]{fontenc}
\usepackage[usenames,dvipsnames]{xcolor}
\usepackage{setspace}
\usepackage[compact]{titlesec}

\usepackage{epstopdf}

\definecolor{cream}{RGB}{222,217,201}

\newcommand{\Duo}{{\sc Duo}}

\newcommand{\cm}{cm\textsuperscript{-1}}

\newcommand{\ai}{{\it ab initio}}

\newcommand{\X}{$X\,{}^{2}\Sigma^{+}$}
\newcommand{\A}{$A\,{}^{2}\Pi$}
\newcommand{\C}{$C\,{}^{2}\Pi$}
\newcommand{\Ap}{$A'\,{}^{2}\Delta$}
\newcommand{\B}{$B\,{}^{2}\Sigma^{+}$}
\newcommand{\D}{$D\,{}^{2}\Sigma^{+}$}
\newcommand{\aq}{$a\,{}^{4}\Pi$}
\newcommand{\onlinecite}[1]{\hspace{-1 ex} \nocite{#1}\citenum{#1}}

\newcommand{\rme}{{\rm e}}

\newcommand{\allstates}{\X, \Ap, \A, \B, \C, \D}

\begin{document}

\title{Spectroscopy of YO from first principles$^\dag$}
\author{Alexander N. Smirnov\textit{$^{a}$},  Victor G. Solomonik\textit{$^{a}$}, Sergei N. Yurchenko$^{\ast}$\textit{$^{b}$} and Jonathan Tennyson\textit{$^{b}$}}

\maketitle

\section*{Abstract}

We report an \ai\ study on the spectroscopy of the open-shell diatomic molecule yttrium oxide, YO. The study
considers the six lowest doublet states, \allstates, and a few higher-lying quartet states using high levels
of electronic structure theory and accurate nuclear motion calculations. The coupled cluster singles,
doubles, and perturbative triples, CCSD(T), and multireference configuration interaction (MRCI) methods are
employed in conjunction with a relativistic pseudopotential on the yttrium atom and a series of
correlation-consistent basis sets ranging in size from triple-$\zeta$ to quintuple-$\zeta$ quality.
Core--valence correlation effects are taken into account and complete basis set limit extrapolation is
performed for CCSD(T). Spin-orbit coupling is included through the use of both MRCI state-interaction with
spin--orbit (SI-SO) approach and four-component relativistic equation-of-motion CCSD calculations. Using the
\ai\ data for bond lengths ranging from 1.0 to 2.5~\AA, we compute 6 potential energy, 12 spin--orbit, 8
electronic angular momentum, 6 electric dipole moment and 12 transition dipole moment (4 parallel and 8
perpendicular) curves which provide a complete description of the spectroscopy of the system of six lowest
doublet states. The \textsc{Duo} nuclear motion program is used to solve the coupled nuclear motion
Schr\"{o}dinger equation for these six electronic states. The spectra of $^{89}$Y$^{16}$O simulated for
different temperatures are compared with several available high resolution experimental studies; good
agreement is found once minor adjustments are made to the electronic excitation energies.


\renewcommand*\rmdefault{bch}\normalfont\upshape
\rmfamily



\footnotetext{\textit{$^{a}$~Department of Physics, Ivanovo State University of Chemistry and Technology,
Ivanovo 153000, Russia}} \footnotetext{\textit{$^{b}$~Department of Physics \& Astronomy, University College
London, London WC1E~6BT, UK }}

\footnotetext{$^{\ast}$~Corresponding author; E-mail: s.yurchenko@ucl.ac.uk }
\footnotetext{\dag~Electronic Supplementary Information (ESI) available. See DOI: 10.1039/b000000x/}

\section{Introduction}\label{sec:intro}

Oxides of transition metals and lanthanides have rich and complex spectra due to the presence of many
low-lying excited electronic states. This complexity poses particular challenges for experimental
\cite{09Bexxxx.VO} and theoretical \cite{jt623} studies. The yttrium oxide, YO, is an example of a rare-earth
oxide whose electronic structure is very difficult to explore. Yttrium is a relatively abundant rare-earth
element both on Earth (the 28th most abundant element \cite{06Cotton.book}) and in space (the second most
abundant rare-earth metal \cite{89AnGrxx}). As a result, the spectrum of YO has been the subject of many
astrophysical observations. In particular, YO has been observed in a variety of spectra of cool stars
including R-Cygni \cite{82Muxxxx.YO}, Pi-Gruis \cite{83Muxxxx.YO}, V838 Mon \cite{07GoBaxx.YO,09KaScTy.YO},
and V4332 Sgr \cite{07GoBaxx.YO}.
The spectrum of YO has been extensively used as a probe to study high temperature materials at the focus of a solar
furnace \cite{05BaCaG1.YO,05BaCaGr.YO,07BaCaG1.YO}.
The
\A$_{1/2}$ electronic state YO has a relatively short life time of 33 ns \cite{77LiPaxx.YO}  with large
diagonal Franck-Condon factors,\cite{83BeGrxx.YO} which makes this molecule well suited for cooling
experiments with the potential in quantum information applications.\cite{13HuYeSt.YO}  Yttrium oxide is one
of the very few molecules that have been laser cooled and trapped in a magneto-optical trap
\cite{15YeHuCo.YO,15CoHuYe.YO,16QuGoJo.YO,18CoDiWu.YO}.


A considerable number of experimental studies have been performed probing the \A\ -- \X,
\cite{77ShNixx.YO,78Lixxxx.YO,79BeBaLu.YO,79LiPaxx.YO,80WiDiZe.YO,82BaMuxx.YO,83BeGrxx.YO,84WiDiZe.YO,88ChPoSt.YO,90StShxx.YO,91DyMuNo.YO,93FrKuRe.YO,93OtGoxx.YO,02BaGrxx.YO,03BaGrxx.YO,05BaCaG1.YO.YO,05BaCaGr.YO,06KoSexx.YO,07BaCaG1.YO,07BaCaGr.YO}
\B\ -- \X,\cite{77ShNixx.YO,79BeBaLu.YO,80BeGrxx.YO,93FrKuRe.YO,05LeMaCh.YO,17ZhZhZh} \Ap\ -- \X,
\citep{76ChGoxx.YO,92SiJaHa.YO,15CoHuYe.YO} and \D\ -- \X\ \cite{17ZhZhZh} bands of YO, as well as its
microwave rotational spectrum \cite{61UhAkxx.YO,86StAlxx.YO,93HoToxx.YO} and its hyperfine
structure.\cite{65KaWexx.YO,86StAlxx.YO,87StAlxx.YO,88ChPoSt.YO,90SuLoFr.YO,99KnKaPe.YO,03StVixx.YO}
Chemiluminescence spectra of YO have also been investigated.\cite{75MaPaxx.YO,77ChGoxx.YO,93FrKuRe.YO} Many
of these spectra were recorded using YO samples which were not in thermodynamic equilibrium, thus, at best,
only providing information on the relative intensities. For YO, relative intensity measurements were carried
out for the \A\ -- \X\ system by \citet{82BaMuxx.YO}. However, the permanent dipole moments of YO in both the
\X\ and \A\ states were measured using the Stark technique.\cite{90StShxx.YO,90SuLoFr.YO,03StVixx.YO}

In case of the absence of direct intensity measurements, measured lifetimes can provide important information
on Einstein A coefficients and hence transition dipole moments.\cite{jt624} The lifetimes of some lower lying
vibrational states of YO in its \A, \B, and \D\ states were measured by \citet{77LiPaxx.YO} and
\citet{17ZhZhZh}.

YO is a strongly bound system. The compilation by \citet{68Gaydon.book} reports its dissociation energy
to be 7.0$\pm$2 eV, while \citet{74AcRaxx} recommended a $D_0$ value of 7.290(87) eV based on mass
spectrometric determinations.

A few theoretical investigations of YO are available in the literature. The most comprehensive one was
carried out by \citet{88LaBaxx.YO} who reported the spectroscopic constants for the lowest five doublet, \X,
\Ap, \A, \B, \C, and fourteen quartet electronic states of YO. The doublets were studied at the
multireference single and double excitation configuration interaction (MRCI) level of theory and, in the case
of the \X, \Ap, and \A\ states, also using the modified coupled-pair functional (MCPF) method. All the
quartet states were considered at the CASSCF level, and that with the lowest energy, reportedly $^4\Phi$, at
the MCPF level as well.  \citet{17ZhZhZh} have recently reported the CASPT2 spectroscopic constants and
excitation energies for a set of lowest doublet states of YO including the \D\ state in addition to the
doublets studied previously by \citet{88LaBaxx.YO}. In all of the previous theoretical studies, only modest
double-$\zeta$~\cite{88LaBaxx.YO} or triple-$\zeta$~\cite{17ZhZhZh} basis sets were employed. RKR curves and
some Franck-Condon factors of YO were computed by \citet{11SrShxx.YO}.

The main objective of the present study is to characterize both the electronic ground state and the plethora
of low-lying excited states of YO with high-level \ai\ methods, and to accurately describe from first
principles the spectroscopy of YO via producing the potential energy curves (PECs) and other data needed to
calculate the rovibronic energies and transition probabilities comprising a so-called line list for this
molecule. The generation of such line lists is a major object of the ExoMol project.\cite{jt528}

Thus far, ExoMol studies of open-shell transition metal (TM) diatomics have struggled due to difficulties in
providing reliable \ai\ starting points.~\cite{jt599,jt623,jt644,jt760} The intrinsic challenge to theory
posed by open-shell systems is associated with several types of problems including spin contamination,
symmetry breaking in the reference function, strong nondynamical electron correlation effects, avoided
crossings between adiabatic potential energy surfaces, etc. (for the discussion, see, e.g.,
Refs.\citenum{03StGaxx,jt632}). In the open-shell TM-containing species, these problems are exacerbated by
stronger relativistic effects than those in the molecules made up of relatively light main group elements,
and greater number of electronic excited states governing the spectroscopic behaviour of a molecule and hence
deserving to be taken into account in a study aimed at accurate description of its spectroscopy. Moreover,
the low-lying electronic states of TM species are commonly degenerate or near-degenerate, which complicates
their theoretical treatment even more. Multireference methods of quantum chemistry best suited for describing
closely spaced electronic states might seem to be the natural choice for studying these systems. However,
most routine multireference methods, such as MRCI, are incapable of properly handling dynamical electron
correlation and therefore do not provide high accuracy description of TM-containing species commonly
featuring strong dynamical correlation effects. Such effects are best treated with single reference coupled
cluster (CC) theory known for its capability to predict highly accurate properties even for molecules with
mild to moderate MR character. Unfortunately, the higher likelihood of severe multireference character in the
ground and/or low-lying electronic excited states of open-shell TM-containing species makes their treatment
by single reference methods very problematic, if possible at all. Particularly this is true for the studies
aimed at a description of the molecular potential energy surfaces over a wide range of geometries. It is
therefore not surprising that the high-level coupled cluster studies on the open-shell TM-containing species,
where a few excited states are treated on an equal footing with the ground state, are very uncommon and only
deal with near-equilibrium regions of these states (see, e.g., Refs.
\citenum{04HoPaYa,05PuPexx,06PaYaSc,07PaYaSc}). Such a study on a manifold of electronic excited states of a
TM-containing diatomic molecule over a wider bond length range has not been reported so far.

It is thus clear that none of routine methods of modern quantum chemistry are entirely satisfactory in all
respects for accurately describing from first principles the spectroscopy of open-shell TM-containing
species. Nevertheless, one can try to solve this challenging task via the so-called composite approach by
which the desired set of molecular properties is obtained using multiple methods of different nature and
sophistication rather than a single method.

In this paper, we have examined efficiency of such an approach taking the example of YO. The PECs for the six
lowest doublet electronic states of this molecule, \allstates, were obtained from the extensive high-level
coupled cluster calculations addressing core--valence correlation and basis set convergence issues, whereas
the spin--orbit curves (SOCs), electronic angular momentum curves (EAMCs), electric dipole moment curves
(DMCs), and transition dipole moment curves (TDMCs) were obtained at the MRCI level of theory. These curves,
with some simple adjustment of the minimum energies of the PECs, are used to solve the coupled nuclear-motion
Schr\"{o}dinger equation with the program \Duo.\cite{jt609} The spectroscopic model and \ai\ curves are
provided as part of the supplementary material. Our open source code \Duo\ can be accessed via
\verb!http://exomol.com/software/!.

\section{Computational details}

\subsection{Ab initio calculations}

Multireference single and double excitation configuration interaction, MRCI,
\cite{88WeKnxx.ai,88KnWexx.ai,92KnWexx.ai} and underlying complete active space self-consistent field,
CASSCF,\cite{85WeKnxx.ai,85KnWexx.ai} calculations were carried out using a relativistic energy-consistent
28-electron core pseudopotential (PP) accompanied with the aug-cc-pVTZ-PP \cite{07PeFiDo.ai} basis set on the
Y atom, and the aug-cc-pVTZ \cite{92KeDuHa.ai} all-electron basis set on the O atom (this combination of sets
is hereafter referred to as aVTZ). To obtain a consistent MRCI data set in the widest possible range of bond
lengths, the state-averaged CASSCF procedure was employed with density matrix averaging over 22 doublet (six
$\Sigma^+$, seven $\Pi$, five $\Delta$, two $\Phi$, and two $\Sigma^-$) and 9 quartet (two $\Sigma^+$, three
$\Pi$, two $\Delta$, one $\Phi$, and one $\Sigma^-$) states, with equal weights for each of the roots. The
active space included 7 electrons distributed in 13 orbitals (6 $a_1$, 3 $b_1$,  3$b_2$, 1 $a_2$) that had
predominantly O 2p and Y 4d, 5s, 5p, and 6s character; all lower energy orbitals were constrained to be
doubly occupied. All valence electrons (4d, 5s Y; 2s, 2p O) were included in the MRCI correlation treatment.

Potential energy, spin--orbit coupling, and dipole moment curves, as well as electronic angular momentum and
transition dipole matrix elements were obtained at the MRCI level for the six lowest doublet states.
Moreover, the potential curves were calculated using the extended multi-state complete active space
second-order perturbation theory,\cite{11ShGyCe.ai} XMS-CASPT2, with the basis sets
aug-cc-pwCVTZ-PP\cite{07PeFiDo.ai} on Y and aug-cc-pwCVTZ\cite{02PeDuxx.ai} on O (henceforth abbreviated as
awCVTZ). In the respective SA-CASSCF calculations, the (7e,13o) active space was employed together with
averaging over the lowest six doublet states. In order to remedy issues pertaining to intruder states, a
level shift of 0.4 and an IPEA (ionisation potential, electron affinity) shift of 0.5 were employed for
XMS-CASPT2.

To calculate the molecular $\Omega$ states and respective spin--orbit curves, we used the spin--orbit -- MRCI
state-interacting approach\cite{00BeScWe.ai}: the spin-coupled eigenstates were obtained by diagonalizing
$H_{\rm es}$ + $H_{\rm SO}$ in a basis of MRCI eigenstates of electrostatic Hamiltonian $H_{\rm es}$. The
matrix elements of $H_{\rm SO}$ were constructed using the one-electron spin--orbit operator accompanying the
yttrium pseudopotential.

Spin--orbit effects were also treated more rigorously in relativistic four-component (4c) all-electron
calculations employing a Gaussian nuclear model and an accurate approximation to the full Dirac--Coulomb
Hamiltonian.\cite{97Visscher.ai} The respective spin-free results were obtained with the spin-free
Hamiltonian of Dyall.\cite{94Dyall.ai} In these calculations, the relativistic TZ-quality basis sets of
Dyall\cite{07Dyall.ai,16Dyall.ai} were used for the Y and O atoms (hereafter referred to as TZ$_D$). The
basis sets were kept uncontracted to provide sufficient flexibility. Electron correlation was taken into
account via the equation-of-motion CCSD (EOM-CCSD) method\cite{18ShSaVi.ai} with the Y outer-core (4s and 4p)
electrons correlated together with the valence electrons. The EOM-EA scheme (adding 1 electron to the closed
shell) was applied with the reference defined by the YO$^+$ cation and the active space comprising 12 spinors
(Y 5s and 4d). For the YO electronic states inaccessible via the EOM-EA procedure, we employed the EOM-IP
scheme (removing 1 electron from the closed shell) with the YO$^-$ (Y 5s$^2$) anion taken as the reference
and an active space composed of 8 spinors (Y 5s and O 2p). The virtual orbital space was truncated by
deleting all virtual spinors with orbital energies larger than 15 a.u. In the relativistic calculations of
dipole moments, a finite-field perturbation scheme was employed by adding the z-dipole moment operator as a
small perturbation to the Hamiltonian. Perturbations with electric field strengths of $\pm$0.0005 a.u. were
applied.

The atomic spin--orbit corrections, $\Delta E_{\rm SO}$, utilized in the calculations of the YO atomization
energy were obtained from the experimental $J$-averaged zero-field splittings of the ground state atomic
terms \cite{NIST}: $\Delta E_{\rm SO}$ = $-$77.975 \cm\ (O) and $-$318.216 \cm\ (Y).

The most sophisticated PEC computations were performed at the coupled cluster singles, doubles, and
perturbative triples, CCSD(T), level of theory\cite{92HaPeWe.ai} with a restricted open-shell  Hartree-Fock
reference and with an allowance for a small amount of spin contamination in the solution of the CCSD
equations, i.e., RHF-UCCSD(T). \cite{93WaGaBa.ai,93KnHaWe.ai} Symmetry equivalencing of the ROHF orbitals was
performed for the degenerate atomic and molecular electronic states. Both valence (4d, 5s Y; 2s, 2p O) and
outer-core (4s, 4p Y; 1s O) electrons were correlated. Scalar relativistic effects were treated with the
yttrium pseudopotential described above. Sequences of aug-cc-pwCV$n$Z-PP\cite{07PeFiDo.ai} ($n$ = T, Q, 5) basis
sets for Y were used in conjunction with the corresponding all-electron basis sets
aug-cc-pwCV$n$Z\cite{02PeDuxx.ai} for the O atom. These combinations of basis sets are denoted below as
awCVTZ, awCVQZ, and awCV5Z, respectively.

For each point in a grid of $r$(Y--O) bond lengths, the CCSD(T) calculated energies were extrapolated to the
complete basis set (CBS) limit. Three extrapolation schemes were employed. First, a two-point extrapolation
of total energies was performed using the formula \cite{96Martin.ai}:
\begin{equation}
\label{eq:2p} E_{\mbox{\scriptsize $n$}} = E_{\mbox{\scriptsize CBS}} +
\frac{A}{(n+1/2)^4} ,
\end{equation}
where $n$ = 4 and 5 for the awCVQZ and awCV5Z basis sets. This scheme is denoted as CBS1. Second, we employed
alternative two-point (Q5) extrapolations of the Hartree--Fock and correlation energy components. These
implied using Eq. \ref{eq:2p} for the correlation part and the Karton and Martin formula\cite{06KaMaxx.ai} for
the HF energy:
\begin{equation}
\label{eq:sep} E_{\mbox{\scriptsize $n$}} = E_{\mbox{\scriptsize CBS}} + A \,
(n+1) \, \exp{ (-9.03 n^{1/2} )};
\end{equation}
this is denoted as CBS2. Third, the CBS estimates were also obtained using the awCVTZ, awCVQZ, and awCV5Z
total energies via the three-parameter, mixed Gaussian/exponential expression \cite{94PeWoDu.ai}:
\begin{equation}
\label{eq:3p} E_{\mbox{\scriptsize $n$}} = E_{\mbox{\scriptsize CBS}} +  A \, \exp{(-(n-1))} + B \,
\exp{ (-(n-1)^2)},
\end{equation}
where $n$ = 3, 4 and 5 for the awCVTZ, awCVQZ and awCV5Z basis sets, respectively. This is denoted as CBS3.

The spectroscopic constants $r_\rme$, $\omega_\rme$, $\omega_\rme$$x_\rme$, and $\alpha_\rme$ of YO were
obtained from a conventional Dunham analysis\cite{32Dunham} using polynomial fits of total energies for bond
lengths in the vicinity of the minimum for a given electronic state.

The CCSD(T) calculations of the equilibrium dipole moments, $\mu_e$, for a few lowest states were
carried out at the corresponding CCSD(T)/CBS1 equilibrium bond lengths. The dipole moments were computed
by numerical differentiation of the total energy in the presence of a weak electric field. Finite
perturbations with electric field strengths of $\pm$0.0025 a.u. were applied. Since hierarchical
sequences of basis sets have been used, the dipole moments were also  extrapolated to the CBS limit using
the three extrapolation schemes described above.

Most of the ab initio calculations were carried out using the MOLPRO electronic structure
package.\cite{MOLPRO2015} The relativistic 4c-EOM-CCSD calculations were done with the use of the DIRAC
program.\cite{Dirac18}

\subsection{Duo calculations}

We use the program \Duo\ \cite{jt609,jt626} to solve the coupled Schr\"{o}dinger equation for 6 lowest
electronic states of YO. \Duo\ is a variational program capable of solving rovibronic problems for a general
(open-shell) diatomic molecule with an arbitrary number of couplings, see, for example,
Refs.\citenum{jt589,jt736,jt759,jt760}. All \ai\ couplings between these 6 states are taken into account as
described below.  The goal of this paper is to provide a qualitative simulation of the electronic spectra of
YO based on the \ai\ curves. We therefore do not attempt a systematic refinement of the \ai\ curves by
fitting to the experiment, which will be the subject of future work. In order to facilitate the comparison
with the experimental data, we, however, perform some shifts of the $T_{\rm e}$ values and simple scaling of
the SOCs (see below).

In \Duo\ calculations, the coupled Schr\"{o}dinger equation was solved on an equidistant grid of 301 bond
lengths $r_i$ ranging from  $r = $ 1.2 to 3 \AA\ using the sinc DVR method. Our \ai\ curves are represented
by sparser and less extended grids (see below). For the bond length values $r_i$ overlapping with the \ai\
ranges, the \ai\ curves were projected onto the denser \Duo\ grid using the cubic spline interpolation. The PECs
outside the \ai\ range were reconstructed using the standard Morse potential form
$$
  f_{\rm PEC}(r) = V_{\rm e} + D_{\rm e} \left( 1-e^{-a (r-r_{\rm e})} \right)^2. \\
$$
For other curves the following function forms were used \citep{jt609}:
\begin{eqnarray}
\nonumber
  f_{\rm TDMC}^{\rm short}(r) &=& A r + B r^2,\\
  \nonumber
  f_{\rm other}^{\rm short}(r) &=& A + B r,
\end{eqnarray}
for the short range and
\begin{eqnarray}
\nonumber
 f_{\rm EAMC}^{\rm long}(r)  &=& A + B r, \\
 \nonumber
 f_{\rm other}^{\rm long}(r)  &=& A/r^2 + B/r^3.
\end{eqnarray}
for the long range, where $A$ and $B$ are stitching parameters.

The vibrational basis set was taken as eigensolutions of the six uncoupled 1D problems for each PEC.  The
corresponding basis set constructed from 6$\times $301 eigenfunctions was then contracted to include 60
lowest (in terms of the vibrational energy) $X$ functions and 20 from each other state (160 in total). These
vibrational basis functions were then combined with the spherical harmonics for the rotational and electronic
spin basis set functions. All calculations were performed for $^{89}$Y$^{16}$O using atomic masses.

This study is the first where a \Duo\ calculation has been performed for a system with avoided crossings
between curves of the same symmetry. The current version of \Duo\ does not allow for non-adiabatic couplings,
and therefore these were ignored in this study. However, despite the expectation that the non-adiabatic
coupling effects should be relatively important in the regions near an avoided curve crossing, as we show
below, our model neglecting these effects is justified by close agreement with the available experimental
spectra.

\section{Results and discussion}

\subsection{Results of ab initio calculations}

\subsubsection{Electronic structure and potential energy curves of the YO molecule}

\begin{figure}
	\includegraphics[width=0.4\textwidth]{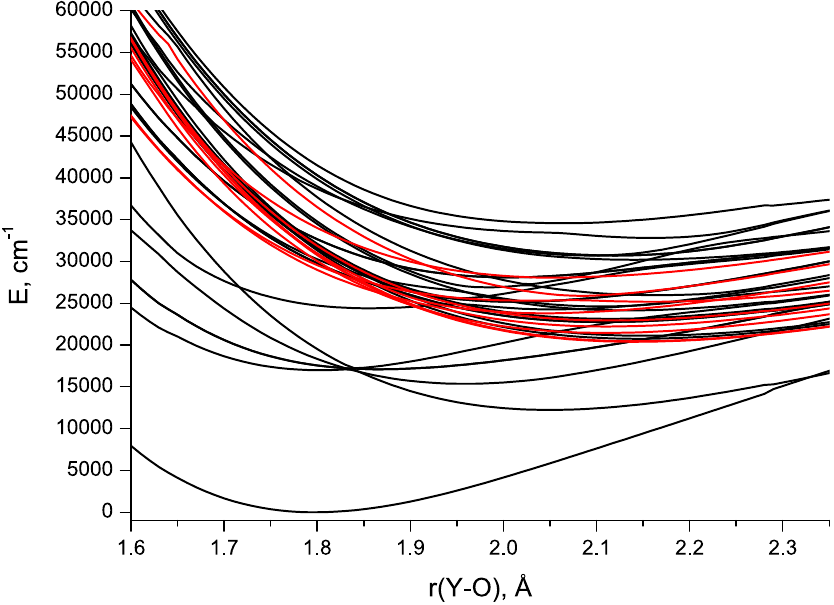}
	\caption{CASSCF/aVTZ spin-free potential energy curves of the
low-lying doublet (black) and quartet (red) states in YO.}
	\label{f:CAS2-4}
\end{figure}

An overview of the CASSCF PECs for all doublet and quartet states included in the SA-CASSCF procedure is
provided in Fig.~\ref{f:CAS2-4}. In the vicinity of the ground state minimum (at $\sim$1.8 \AA), the lowest
six CASSCF states are doublets, whereas the quartet states lie at $\sim$30000 \cm\ above the ground state.

\begin{figure}[H]
  \includegraphics[width=0.4\textwidth]{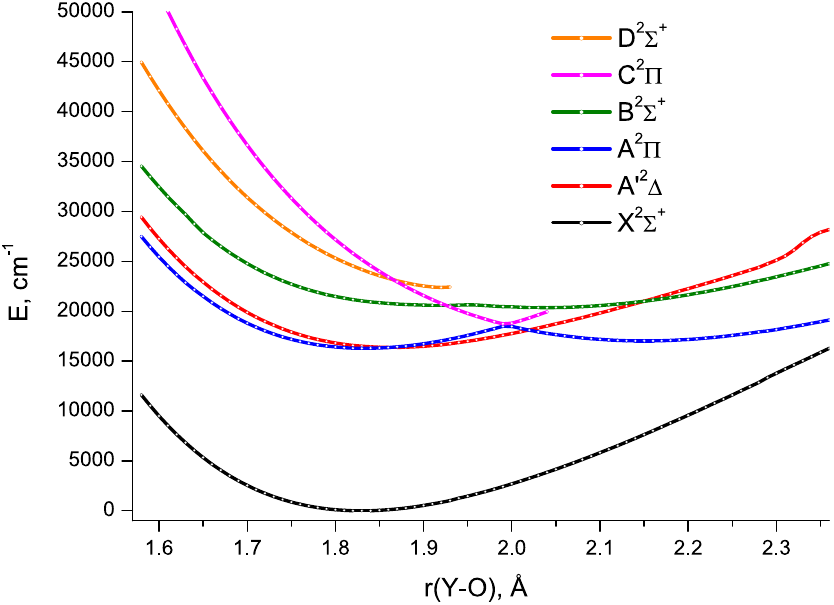}
	\caption{MRCI/aVTZ spin-free potential energy curves of YO.}
  \label{f:PECsCI}
\end{figure}

\begin{figure}[H]
  \includegraphics[width=0.4\textwidth]{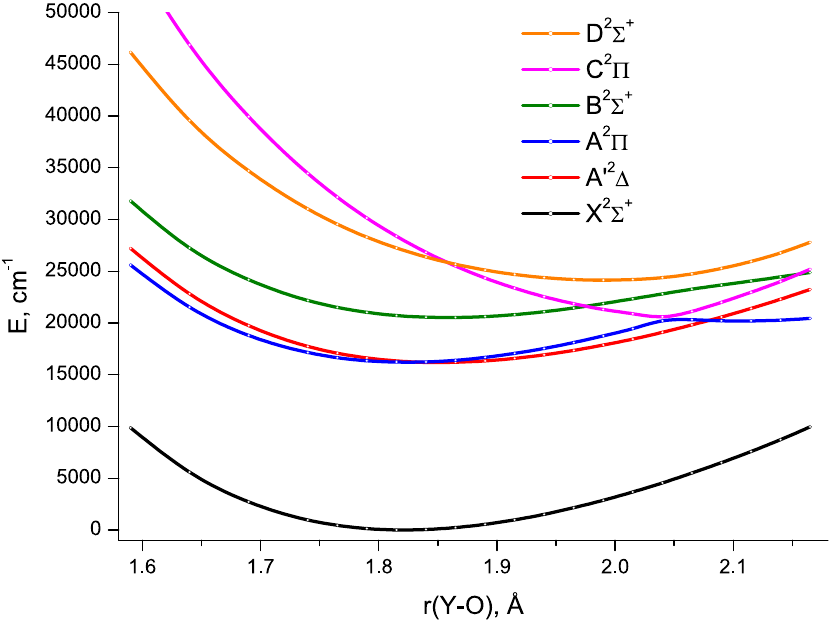}
	\caption{XMS-CASPT2/awCVTZ spin-free potential energy curves of YO.}
  \label{f:XMS-CAS}
\end{figure}

\begin{figure}[H]
  \includegraphics[width=0.4\textwidth]{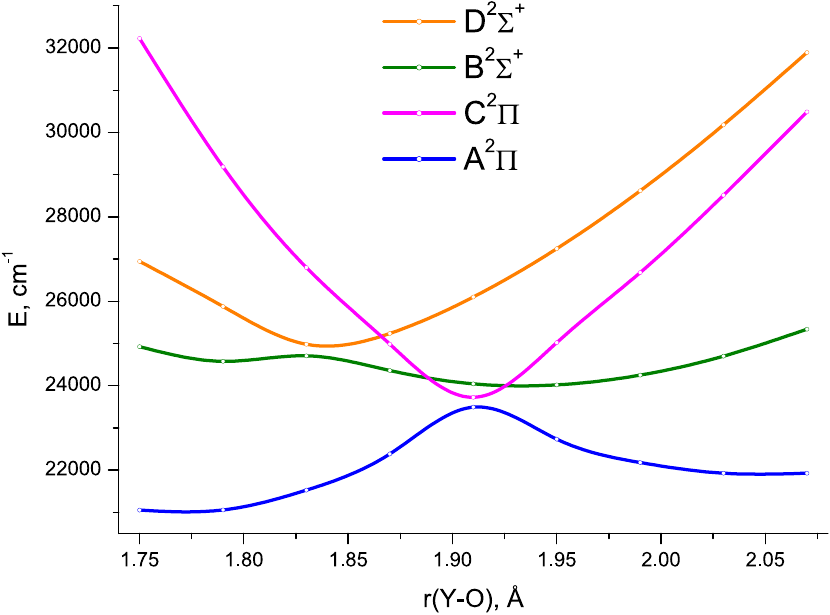}
	\caption{The avoided crossing regions of the EOM-IP-CCSD/TZ$_D$ spin-free potential energy curves for the \A, \B, \C, and \D\ electronic states of YO.}
  \label{f:eom-sf}
\end{figure}

The lowest six doublet MRCI PECs (\allstates) are shown in Fig.~\ref{f:PECsCI}. For the states \X, \Ap, \A,
and \B, the PECs were obtained in the full bond length range amenable to the underlying CASSCF procedure,
$1.58 \leq r \leq 2.36$ \AA, while the \C\ and \D\ curves were calculated through $r = 2.04$ \AA\ and $1.93$
\AA, respectively. Extending the MRCI curves for the two upper states beyond those distances would require
requesting a greater number of states (while exceeding 4 in a single irreducible representation with the
chosen active space would make the MRCI computation unfeasible on the hardware used) or selecting the order
of the states in the initial internal CI (e.g., 1, 2, 3, 5 rather than 1, 2, 3, 4), leaving some of them out
in each MRCI point. This appeared to affect the smoothness of the resulting PECs. Therefore, we refrained
from further pursuing the computations with the same number of MRCI roots in the entire bond length range and
simply reduced the number of requested states for longer internuclear distances. For all six doublet states,
the XMS-CASPT2 PECs could be obtained in the range $1.59 \leq r \leq 2.165$ \AA, Fig.~\ref{f:XMS-CAS}; at
larger bond lengths the underlying CASSCF procedure failed to converge. As can be seen from
Figs.~\ref{f:PECsCI} and \ref{f:XMS-CAS}, the \A\ and \C\ curves feature an avoided crossing at bond lengths
around 2~\AA, as do the \B\ and \D\ curves in approximately the same region, see Fig.~\ref{f:XMS-CAS}. The
avoided crossings of both pairs of curves are also seen in the EOM-IP-CCSD calculated PECs,
Fig.~\ref{f:eom-sf}, albeit at shorter distances (1.8 -- 1.9 \AA).

In order to provide deeper insight into the electronic structure of YO, we have analyzed the dominant
configurations (configuration state functions) in the MRCI wave functions for the lowest electronic states,
Table~\ref{t:wtsd}, and the leading atomic orbital contributions in the respective molecular orbitals,
Table~\ref{t:molorb}.

The analysis indicates that the \X\ ground state consists mainly of
...10$\sigma^2$11$\sigma^2$5$\pi^4$12$\sigma^1$ electron configuration. The principal contribution to the
singly occupied 12$\sigma$ MO comes from the yttrium 5s atomic orbital, with an admixture of the 5p AO
particularly noticeable at longer internuclear distances. The three lowest active MOs, 10$\sigma$,
11$\sigma$, and 5$\pi_{+(-)}$, primarily consist of the oxygen 2s and 2p orbitals whose contributions
increase with the bond stretching. Therefore, the bonding in the \X\ YO ground state can be roughly described
as ionic, Y$^{2+}$O$^{2-}$, however, with significant covalent character mainly owing to an appreciable
participation of the yttrium 4p$_\sigma$ AO in the 10$\sigma$ MO. Upon the Y--O bond stretching, there is a
rapid decrease in the 4p contribution (see Table~\ref{t:molorb}) reducing the covalency and resulting in a
concomitant increase in the magnitude of the electric dipole moment in the YO ground state (see below).

The first excited state, \Ap, mainly consists of ...10$\sigma^2$11$\sigma^2$5$\pi^4$2$\delta^1$ configuration
with the 2$\delta$ MO clearly assigned to the Y 4d$_\delta$ orbital. At shorter bond lengths this state can
be reasonably described by the single electron excitation from the ground state, 5s$^1$ $\rightarrow$
4d$_\delta$$^1$. Upon bond elongation, the weight of the main configuration gradually decreases, approaching
$\sim$50\% at bond lengths of about 2.2 \AA, whereas at r > $\sim$2 \AA, a few other large-weight
configurations emerge, e.g., the three-open-shell
...10$\sigma^2$11$\sigma$$^\alpha$12$\sigma$$^\beta$5$\pi^4$2$\delta$$^\alpha$ configuration (with a weight
of 22\% at 2.19 \AA) that corresponds to the Y$^+$O$^-$ bonding (Y 5s$^1$4d$^1$, O 2p$^5$).

At bond lengths up to $\sim$1.85 \AA, the principal configurations of the \A\ and \B\ excited states,
...10$\sigma^2$11$\sigma^2$5$\pi^4$6$\pi^1$ and ...10$\sigma^2$11$\sigma^2$5$\pi^4$13$\sigma^1$,
respectively, include the 6$\pi$ and 13$\sigma$ singly occupied  MOs mainly composed of the yttrium 5p$_\pi$
and 5p$_\sigma$ atomic orbitals, respectively, yet with a significant admixture of the 4d AOs. Therefore,
these states can be viewed as arising from 5s $\rightarrow$ 5p$_\pi$ and 5s $\rightarrow$ 5p$_\sigma$ atomic
electron promotions. In the same range of bond length values, the \C\ and \D\ upper lying states consist
mainly of ...10$\sigma^2$11$\sigma^2$12$\sigma^2$5$\pi^3$ and ...10$\sigma^2$11$\sigma^1$12$\sigma^2$5$\pi^4$
electron configurations, respectively. The YO bonding in the \C\ and \D\ states is hence well described by
the Y$^+$O$^-$ scheme consistent with the Y 5s$^2$, O 2p$^5$ electron configuration.

As the Y--O distance approaches the avoided crossing point from below, the principal configurations of the
\A\ and \B\ states change to those specified above for the \C\ and \D\ states, respectively, while, vice
versa, the principal configurations of the \C\ and \D\ states turn into those being the main ones for the \A\
and \B\ states at shorter bond lengths. This alteration of main configurations describes an oxygen-to-metal
charge back-transfer, Y$^{2+}$O$^{2-}$ $\rightarrow$ Y$^+$O$^-$, in the \A\ and \B\ states of YO upon the
Y--O bond stretching through the avoided crossing region of bond lengths.

Specifically, the avoided crossing point, $r_{ac}$, chosen to be the point of closest approach of two curves,
for the \A\ and \C\ states amounts to 2.046~\AA\ (XMS-CASPT2) and 1.994~\AA\ (MRCI), with the energy gap,
$\Delta$$E_{ac}$, of 366 \cm\ and 243 \cm, respectively. For the XMS-CASPT2 \B\ and \D\ curves, $r_{ac}$ =
2.064~\AA\ and $\Delta$$E_{ac}$ = 1484 \cm. Notably, the principal configuration interchange between the \B\
and \D\ states occurs at a slightly shorter internuclear distance: $\sim$1.95~\AA\ (XMS-CASPT2),
$\sim$1.92~\AA\ (MRCI). At the EOM-IP-CCSD level, the avoided crossing characteristics are: $r_{ac}$ =
1.911~\AA, $\Delta$$E_{ac}$ = 231 \cm\ for \A\ and \C\ curves, and $r_{ac}$ = 1.832~\AA, $\Delta$$E_{ac}$ =
272 \cm\ for \B\ and \D\ curves.

The CCSD(T) calculations were carried out for the six lowest doublet states. The reference configurations for
each state were as follows:

\X\  ...10$\sigma^2$11$\sigma^2$5$\pi^4$12$\sigma^1$

\Ap\  ...10$\sigma^2$11$\sigma^2$5$\pi^4$2$\delta^1$

\A\  ...10$\sigma^2$11$\sigma^2$5$\pi^4$6$\pi^1$

\B\  ...10$\sigma^2$11$\sigma^2$5$\pi^4$13$\sigma^1$

\C\  ...10$\sigma^2$11$\sigma^2$12$\sigma^2$5$\pi^3$

\D\  ...10$\sigma^2$11$\sigma^1$12$\sigma^2$5$\pi^4$

\begin{figure}[H]
  \includegraphics[width=0.4\textwidth]{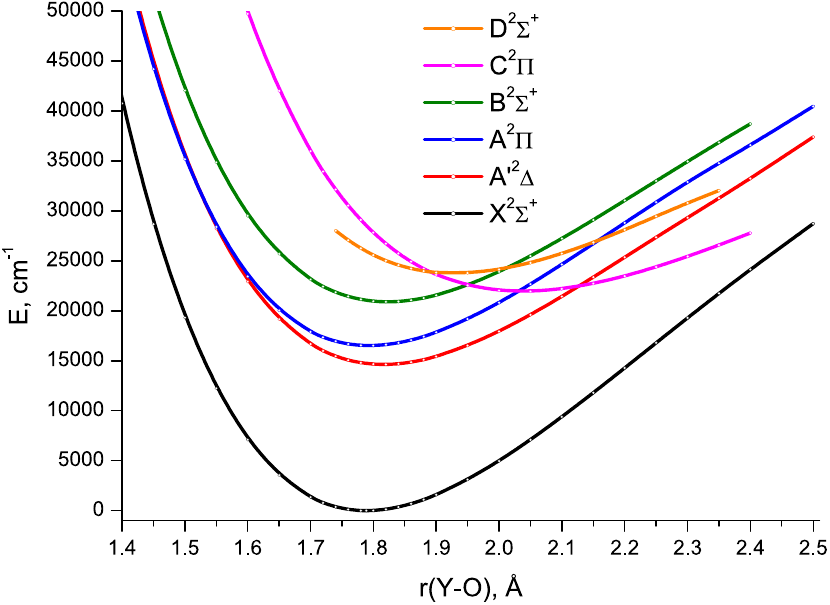}
  \caption{CCSD(T)/CBS1 spin-free potential energy curves of YO.}
  \label{f:PECsCC}
\end{figure}

The CCSD(T) energies were obtained in the ranges: $1.0 \leq r \leq 2.5$ \AA\ for \X, \Ap\ and \A;  $1.0 \leq
r \leq 2.4$ \AA\ for \B; $1.4 \leq r \leq 2.4$ \AA\ for \C; $1.74 \leq r \leq 2.35$ \AA\ for \D. At longer
distances (as well as shorter ones for \C\ and \D), the coupled-cluster calculations failed due to severe
CCSD convergence problems. For the \X, \Ap, \A, and \B states, the distances shorter than 1.0 \AA\ were not
considered because relative energies of excited states already exceed 350000 \cm\ at this point. The
respective CCSD(T)/CBS1 PECs are shown in Fig.~\ref{f:PECsCC}. Since single reference methods are not
suitable for describing avoided crossings, e.g., those between the PECs of the \A\ and \C, and \B\ and \D\
states, the CCSD(T) calculated PECs can be viewed as corresponding to the diabatic presentation of these
states. It is clearly seen that the CCSD(T) \A\ and \C\ curves cross each other at 2.031~\AA, as do the \B\
and \D\ ones at 2.013~\AA.

\begin{table*}
\caption{Weights of the leading configurations (configuration state functions) in the low-lying electronic
states of YO derived from analyzing the MRCI/aVTZ wave functions at bond lengths of 1.79, 2.04 and 2.19 \AA.
Contributions of 2$\%$ and higher are shown.} \label{t:wtsd} \centering
\begin{threeparttable}
\begin{tabular}{lccccccccccrrr}
\hline
\hline
& \multicolumn{10}{c}{Configurations\tnote{a}} & \multicolumn{3}{c}{Weights, $\%$} \\
State & 10$\sigma$ & 11$\sigma$ & 12$\sigma$ & 13$\sigma$ & 5$\pi_{+}$ & 5$\pi_{-}$ & 6$\pi_{+}$ & 6$\pi_{-}$ & 2$\delta_{+}$ & 2$\delta_{-}$ & 1.79 \AA & 2.04 \AA & 2.19 \AA \\
\hline
\X  & 2 & 2 & +   & 0   & 2   & 2   & 0   & 0   & 0 & 0 & 79.2 & 71.4 & 63.3 \\
    & 2 & + & +   & $-$ & 2   & 2   & 0   & 0   & 0 & 0 &  2.1 &  3.1 &  3.5 \\
    & 2 & + & $-$ & +   & 2   & 2   & 0   & 0   & 0 & 0 &      &  2.3 &  2.8 \\
    & 2 & 2 & +   & 0   & 2   & +   & 0   & $-$ & 0 & 0 &      &      &  3.6 \\
    & 2 & 2 & +   & 0   & +   & 2   & $-$ & 0   & 0 & 0 &      &      &  3.6 \\
    & 2 & 2 & +   & 0   & $-$ & 2   & +   & 0   & 0 & 0 &      &      &  2.2 \\
    & 2 & 2 & +   & 0   & 2   & $-$ & 0   & +   & 0 & 0 &      &      &  2.2 \\
\hline
\Ap\ ($a_1$) & 2 & 2 & 0   & 0   & 2 & 2 & 0 & 0 & + & 0 & 78.9 & 67.3 & 53.1 \\
    & 2 & + & $-$ & 0   & 2 & 2 & 0 & 0 & + & 0 &      &  8.1 & 21.6 \\
    & 2 & + & 0   & $-$ & 2 & 2 & 0 & 0 & + & 0 &  4.8 &  4.2 &  4.1 \\
\hline
\A\ ($b_1$) & 2 & 2 & 0 & 0   & 2 & 2 & + & 0 & 0 & 0 & 79.9 &      &      \\
    & 2 & + & 0 & $-$ & 2 & 2 & + & 0 & 0 & 0 &  3.2 &      &      \\
    & 2 & 2 & 2 & 0   & + & 2 & 0 & 0 & 0 & 0 &      & 82.6 & 82.3 \\
    & 2 & 2 & 0 & 0   & + & 2 & 2 & 0 & 0 & 0 &      &  2.4 &  2.1 \\
    & 2 & 2 & 0 & 0   & + & 2 & 0 & 2 & 0 & 0 &      &  2.1 &      \\
\hline
\B  & 2 & 2 & 0 & + & 2 & 2 & 0 & 0 & 0 & 0 & 73.5 &      &      \\
    & 2 & + & 2 & 0 & 2 & 2 & 0 & 0 & 0 & 0 &  4.5 & 80.2 & 78.9 \\
    & 2 & + & 0 & 2 & 2 & 2 & 0 & 0 & 0 & 0 &  2.1 &      &      \\
    & 2 & + & 0 & 0 & 2 & 2 & 2 & 0 & 0 & 0 &      &  2.3 &  2.0  \\
    & 2 & + & 0 & 0 & 2 & 2 & 0 & 2 & 0 & 0 &      &  2.3 &  2.0  \\
\hline
\C\ ($b_1$) & 2 & 2 & 2   & 0   & + & 2 & 0 & 0 & 0 & 0 & 83.2 &      & \\
    & 2 & 2 & 0   & 0   & + & 2 & 2 & 0 & 0 & 0 &  2.8 &      & \\
    & 2 & 2 & 0   & 0   & + & 2 & 0 & 2 & 0 & 0 &  2.3 &      & \\
    & 2 & 2 & 0   & 0   & 2 & 2 & + & 0 & 0 & 0 &      & 68.7 & \\
    & 2 & + & $-$ & 0   & 2 & 2 & + & 0 & 0 & 0 &      &  4.3 & \\
    & 2 & + & 0   & $-$ & 2 & 2 & + & 0 & 0 & 0 &      &  3.3 & \\
\hline
\D  & 2 & + & 2 & 0 & 2 & 2 & 0 & 0 & 0 & 0 & 79.1 & & \\
    & 2 & 2 & 0 & + & 2 & 2 & 0 & 0 & 0 & 0 &  4.0 & & \\
    & 2 & + & 0 & 0 & 2 & 2 & 2 & 0 & 0 & 0 &  2.6 & & \\
    & 2 & + & 0 & 0 & 2 & 2 & 0 & 2 & 0 & 0 &  2.6 & & \\
\hline
\hline
\end{tabular}
\begin{tablenotes}
\small
\item[a] The orbital names $\pi_{+}$, $\pi_{-}$, $\delta_{+}$, and $\delta_{-}$ indicate $\pi(b_1)$,
    $\pi(b_2)$, $\delta(a_1)$, and $\delta(a_2)$ orbitals, respectively. The MO occupancies represented by 2,
    0 and + or $-$ denote double, zero, and single occupancies with the total spin raised or lowered by 1/2.
\end{tablenotes}
\end{threeparttable}
\end{table*}

\begin{table*}
\caption{Analysis of the YO molecular orbitals in terms of leading atomic orbital contributions (above
10$\%$) at bond lengths of 1.79, 2.04 and 2.19 \AA.} \label{t:molorb} \centering
\begin{tabular}{lccc}
\hline \hline
 MO & 1.79 \AA & 2.04 \AA & 2.19 \AA \\
\hline
10$\sigma$ & 52\% 2s O + 43\% 4p$\sigma$ Y & 71\% 2s O + 27\% 4p$\sigma$ Y & 84\% 2s O + 15\% 4p$\sigma$ Y \\
11$\sigma$ & 63\% 2p$\sigma$ O + 10\% 4d$\sigma$ Y & 72\% 2p$\sigma$ O & 77\% 2p$\sigma$ O \\
12$\sigma$ & 86\% 5s Y & 82\% 5s Y + 12\% 5p$\sigma$ Y & 81\% 5s Y + 13\% 5p$\sigma$ Y \\
13$\sigma$ & 50\% 5p$\sigma$ Y + 36\% 4d$\sigma$ Y & 39\% 5p$\sigma$ Y + 53\% 4d$\sigma$ Y & 34\% 5p$\sigma$ Y + 60\% 4d$\sigma$ Y \\
5$\pi_{+(-)}$ & 94\% 2p$\pi$ O & 95\% 2p$\pi$ O & 96\% 2p$\pi$ O \\
6$\pi_{+(-)}$ & 67\% 5p$\pi$ Y + 32\% 4d$\pi$ Y & 46\% 5p$\pi$ Y + 52\% 4d$\pi$ Y & 34\% 5p$\pi$ Y + 63\% 4d$\pi$ Y \\
2$\delta_{+(-)}$ & 100\% 4d$\delta$ Y & 100\% 4d$\delta$ Y & 99\% 4d$\delta$ Y \\
\hline \hline
\end{tabular}
\begin{tablenotes}
\small
\item[a] See footnote to Table~\ref{t:wtsd} for designations.
\end{tablenotes}
\end{table*}

\subsubsection{Spectroscopic constants and excitation energies}

Table~\ref{t:sfree} summarizes the optimized bond lengths $r_\rme$, harmonic vibrational frequencies
$\omega_\rme$, equilibrium dipole moments $\mu_\rme$, and adiabatic excitation energies $T_\rme$ of the
low-lying doublet states calculated at the XMS-CASPT2, MRCI and EOM-CCSD levels as well as the results from
earlier theoretical studies.\cite{88LaBaxx.YO,17ZhZhZh} Our data in the column entitled "\C" were obtained
from the second minimum in the adiabatic PEC of the \A\ state, i.e., they can be ascribed to the diabatic
representation of the \A\ and \C\ states.

The results of our EOM-CCSD calculations indicate that the anionic reference is less suitable for describing
YO than the cationic one. In Table~\ref{t:sfree}, more accurate cationic-reference EOM-EA-CCSD spectroscopic
constants are listed for all states except for the \C\ and \D\ ones which are not accessible via the electron
attachment procedure and therefore were described at the EOM-CCSD level only via EOM-IP.

The results given in Table~\ref{t:sfree} are obviously inferior to those obtained from high-level CCSD(T)
calculations including core--valence correlation and extrapolation to the CBS limit. The CCSD(T) results are
collected in Table~\ref{t:prop_CC} together with the experimental data available to date. The spread in the
CCSD(T)/CBS results from different CBS extrapolation schemes serve as a rough estimate of the uncertainty in
extrapolation. The good agreement between the CBS estimates and experimentally determined spectroscopic
properties of the \X, \Ap, \A, and \B\ electronic states demonstrates the high accuracy in the CCSD(T)/CBS
PECs of these states for bond lengths in the vicinity of the PECs minima, and is indicative of the mild MR
character of the respective electronic wave functions.

\begin{table*}
\caption{ Theoretical spin-free equilibrium constants of YO in its low-lying doublet states: adiabatic
excitation energies, $T_\rme$ -- \cm, bond lengths, $r_\rme$ -- \AA, vibrational frequencies, $\omega_\rme$
-- \cm, and dipole moments $\mu_\rme$ -- D. The relevant experimental data are listed in
Table~\ref{t:prop_CC}.} \label{t:sfree} \centering
\begin{threeparttable}
\begin{tabular}{llcccccc}
\hline
\hline
 & & \X\ & \Ap\ & \A\ & \B\ & \C\ & \D\ \\
\hline
$T_\rme$& XMS-CASPT2/awCVTZ    & 0 & 16183 & 16210 & 20521 & 20198 & 24139  \\
& MRCI/aVTZ      & 0 & 16370 & 16287 & & 17029 & \\
& EOM-CCSD/TZ$_D$\tnote{a} & 0 & 16096 & 16817 & 21654 & 21896 & 23995 \\
& MCPF\cite{88LaBaxx.YO} & 0 & 15288 & 15728 & & & \\
& MRCI\cite{88LaBaxx.YO} & 0 & 15853 & 15655 & 20039 & 20743 &  \\
& CASPT2\cite{17ZhZhZh} & 0 & 15650 & 17340 & 20570 & 21860 & 23800 \\
\hline
$r_\rme$& XMS-CASPT2/awCVTZ &  1.822 & 1.851 & 1.824 &  1.858 & 2.107 & 1.991 \\
& MRCI/aVTZ    & 1.830 & 1.864 & 1.833 & & 2.149   & \\
& EOM-CCSD/TZ$_D$\tnote{a} & 1.787 & 1.811 & 1.789 & 1.818	& 2.049 & 1.934 \\
& MCPF\cite{88LaBaxx.YO} & 1.811 & 1.842 & 1.813 & & & \\
& MRCI\cite{88LaBaxx.YO} & 1.814 & 1.838 & 1.817 & 1.842 & 2.073 & \\
& CASPT2\cite{17ZhZhZh} & 1.79\ & 1.82 & 1.77 & 1.84 & 1.97 & 1.91 \\
\hline
$\omega_\rme$& XMS-CASPT2/awCVTZ  & 796 & 726 & 763 & 696 & 592 & 696 \\
& MRCI/aVTZ     & 777 & 693 & 750 & & 542  & \\
& EOM-CCSD/TZ$_D$\tnote{a} & 881 & 822 & 847 & 804 & 606 & 648 \\
& MCPF\cite{88LaBaxx.YO} & 855 & 785 & 832 & & & \\
& MRCI\cite{88LaBaxx.YO} & 866 & 801 & 834 & 789 & 638 & \\
\hline
$\mu_\rme$ & MRCI/aVTZ\tnote{b}  & 4.410 & 7.871 & 4.343 & 2.971 & 1.329 &  \\
& EOM-EA-CCSD/TZ$_D$\tnote{c} & 4.905 & 7.867 & 4.147 & 2.164 &  &  \\
& MCPF\cite{88LaBaxx.YO}  & 3.976 & 7.493 & 3.244 &  &  &  \\
\hline
\hline
\end{tabular}
\begin{tablenotes}
\small
\item[a] EOM-IP for \C\ and \D, EOM-EA for the remaining states.
\item[b] calculated for each electronic state at the respective CCSD(T)/CBS1 equilibrium bond length.
\item[c] calculated at a bond length of 1.7932 \AA.
\end{tablenotes}
\end{threeparttable}
\end{table*}

{\setlength{\extrarowheight}{-5pt}
\begin{table*}
  \caption{CCSD(T) spin-free equilibrium constants of YO in its low-lying electronic states:
The dissociation energy $D_0$ (eV) referring to the ground electronic state \X, the excitation energies
$T_\rme$ (\cm) of the \Ap, \A, \B, \C\, \D, and \aq\ states, bond length $r_\rme$ (\AA), spectroscopic
constants $\omega_\rme$ (\cm), $\omega_\rme$$x_\rme$ (\cm) and $\alpha_\rme$ (\cm), and dipole moment
$\mu_\rme$ (D).}
  \label{t:prop_CC}
  \centering
\begin{tabular}{lllllllll}
  \hline
  \hline
  &     & \X\ & \Ap\ & \A\ & \B\ & \C\ & \D\ & \aq\ \\
  \hline
$D_0$, $T_\rme$ & awCVTZ    & 7.060      &  &  &  & & & 28924 \\
& awCVQZ    & 7.207      & 14809 & 16555 & 20893 & 21423 & 23261 & 29296 \\
& awCV5Z    & 7.260      & 14712 & 16538 & 20898 & 21700 & 23528 & 29465 \\
& CBS1  & 7.298      & 14633 & 16526 & 20901 & 21925 & 23745 & 29603 \\
& CBS2  & 7.298      & 14629 & 16525 & 20897 & 21917 & 23741 & 29592 \\
& CBS3  & 7.289      &       &       &       &       &       & 29564 \\
& expt. & 7.290(87)\cite{74AcRaxx}  & 14701\cite{76ChGoxx.YO} & 16530\cite{80BeGrxx.YO} & 20793\cite{80BeGrxx.YO} &       & 23972\cite{17ZhZhZh} \\
\hline
$r_\rme$ & awCVTZ    & 1.7978      &  &  &  & & & 2.0902 \\
& awCVQZ    & 1.7927 & 1.8201 & 1.7971 & 1.8268 & 2.0408 & 1.9345 & 2.0841 \\
& awCV5Z    & 1.7905 & 1.8177 & 1.7950 & 1.8244 & 2.0384 & 1.9323 & 2.0817 \\
& CBS1  & 1.7887 & 1.8157 & 1.7932 & 1.8225 & 2.0365 & 1.9306 & 2.0797 \\
& CBS2  & 1.7890 & 1.8160 & 1.7935 & 1.8228 & 2.0368 & 1.9308 & 2.0799 \\
& CBS3  & 1.7892 &        &        &        &        &        & 2.0802 \\
& expt. & 1.7882\cite{07BaCaG1.YO} & 1.817\cite{92SiJaHa.YO}  & 1.7931\cite{83BeGrxx.YO} & 1.8252\cite{80BeGrxx.YO} &     &   &  \\
\hline
$\omega_\rme$ & awCVTZ    & 855.2 &  &  &  & & & 546.0 \\
& awCVQZ    & 861.4 & 794.0 & 822.5 & 780.6    & 601.8 & 661.2 & 550.6 \\
& awCV5Z    & 864.2 & 797.1 & 825.1 & 783.2    & 603.3 & 662.3 & 552.1 \\
& CBS1  & 866.5 & 799.7 & 827.1 & 785.3    & 604.6 & 663.1 & 553.3 \\
& CBS2  & 866.2 & 799.4 & 826.8 & 785.1    & 604.6 & 663.1 & 553.3 \\
& CBS3  & 865.8 &       &       &          &       &       & 552.9 \\
& expt. & 861.5\cite{07BaCaG1.YO} & 794\cite{77ChGoxx.YO}   & 820\cite{80BeGrxx.YO}   & 765\cite{79BeBaLu.YO}  &     &  &  \\
&       & & & & 759\cite{80BeGrxx.YO} & & &  \\
\hline
$\omega_\rme$$x_\rme$ & awCVTZ    & 2.79 &  &  &  & & & 2.52 \\
& awCVQZ    & 2.78 & 3.06 & 3.17       & 2.94 & 2.58 & 2.60 & 2.53\\
& awCV5Z    & 2.79 & 3.05 & 3.18       & 2.98 & 2.57 & 2.60 & 2.57\\
& CBS1  & 2.79 & 3.04 & 3.19       & 3.01 & 2.57 & 2.61 & 2.60\\
& CBS2  & 2.79 & 3.03 & 3.18       & 3.01 & 2.57 & 2.61 & 2.60 \\
& CBS3  & 2.79 &      &            &      &      &      & 2.59\\
& expt. & 2.84\cite{07BaCaG1.YO} & 3.23\cite{76ChGoxx.YO} & 3.15\cite{07BaCaG1.YO} & 3.97\cite{80BeGrxx.YO} &     &   & \\
& & & & 3.35\cite{83BeGrxx.YO} & & & \\
\hline
$\alpha_\rme$$\cdot${10}$^3$ & awCVTZ    & 1.70 &  &  &  & &  & 1.83 \\
& awCVQZ    & 1.68 & 1.85 & 1.90 & 1.86 & 1.79 & 1.85 & 1.83 \\
& awCV5Z    & 1.68 & 1.85 & 1.91 & 1.87 & 1.80 & 1.86 & 1.83 \\
& CBS1  & 1.68 & 1.85 & 1.91 & 1.87 & 1.80 & 1.87 & 1.83 \\
& CBS2  & 1.68 & 1.84 & 1.91 & 1.87 & 1.80 & 1.87 & 1.83 \\
& CBS3  & 1.68 &      &      &      &      &      & 1.83 \\
& expt. & 1.73\cite{07BaCaG1.YO} & 1.7\cite{76ChGoxx.YO}  & 2.01\cite{83BeGrxx.YO} & 2.49\cite{80BeGrxx.YO} &  &  &     \\
\hline
$\mu_\rme$ & awCVTZ    & 4.615 & 7.595 & 3.711 & 1.749 & 2.059 & 1.256 & 3.605 \\
& awCVQZ    & 4.614 & 7.620 & 3.724 & 1.764 & 2.082 & 1.275 & 3.615 \\
& awCV5Z    & 4.611 & 7.626 & 3.728 & 1.770 & 2.090 & 1.281 & 3.618 \\
& CBS1  & 4.609 & 7.630 & 3.730 & 1.775 & 2.097 & 1.287 & 3.621 \\
& CBS2  & 4.609 & 7.630 & 3.731 & 1.777 & 2.097 & 1.287 & 3.621 \\
& CBS3  & 4.609 & 7.629 & 3.729 & 1.774 & 2.095 & 1.285 & 3.620 \\
& expt. & 4.45(7)\cite{90StShxx.YO}  & & 3.68(2)\cite{90StShxx.YO} & & & & \\
& & 4.524(7)\cite{90SuLoFr.YO} & & & & & & \\
\hline
\hline
\end{tabular}
\end{table*}

\begin{figure}[H]
  \includegraphics[width=0.4\textwidth]{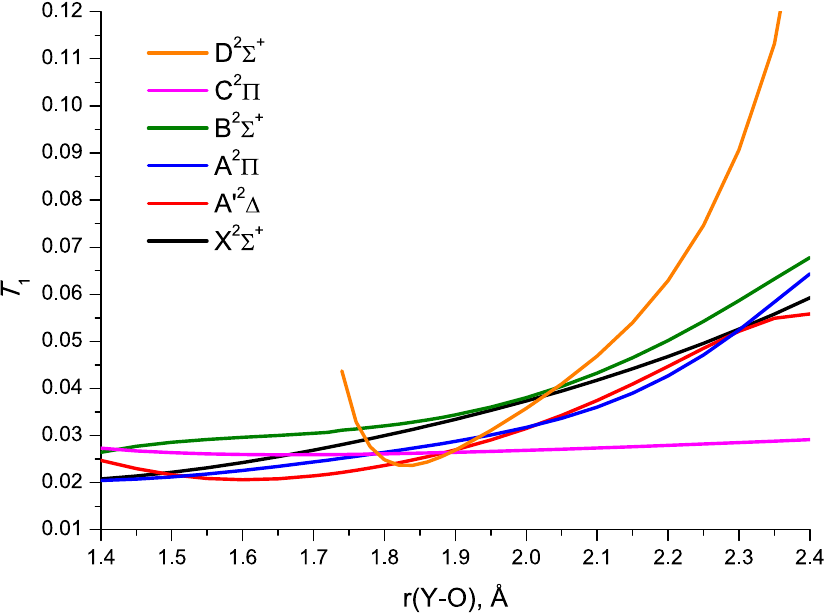}
 \caption{The CCSD/awCV5Z calculated $T_1$ diagnostics of YO}
 \label{f:T1}
\end{figure}

An insight into the reliability of the CCSD(T) PECs over the entire bond length range explored, and for all
electronic states considered, including those not yet characterized experimentally, can be provided by using
the MR diagnostic criteria commonly employed to identify the suitability of single reference
wavefunction-based methods: $T_1$,\cite{89LeTaxx.method} the Frobenius norm of the coupled cluster amplitude
vector related to single excitations, and $D_1$,\cite{98JaNixx.ai} the matrix norm of the coupled cluster
amplitude vector arising from coupled cluster calculations. The utility of different MR diagnostics has been
examined in earlier studies\cite{12JiDeWi.ai,15WaMaWi.ai} on various 3d and 4d TM species. The following
criteria have been proposed~\cite{15WaMaWi.ai} as a gauge for the latter to predict the possible need to
employ multireference wavefunction-based methods while describing energetic and spectroscopic molecular
properties: $T_1$ $\geq$ 0.045, $D_1$ $\geq$ 0.120, $\%$TAE[(T)] $\geq$ 10$\%$. The symbol $\%$TAE[(T)]
denotes here the percent total atomization energy corresponding to a relationship between energies determined
with CCSD and CCSD(T) calculations\cite{06KaRaMa.ai,11KaDaMa.ai}. Obviously, the $\%$TAE[(T)] diagnostic is
applicable for judging the SR/MR character of the electronic ground state only. For the YO molecule, the
CCSD/awCV5Z calculated $\%$TAE[(T)] of 5.6$\%$ is well below the proposed MR threshold. This fact provides
further evidence for single reference character of the \X\ wavefunction in the near-equilibrium region of
Y--O bond lengths.

\begin{figure}[H]
  \includegraphics[width=0.4\textwidth]{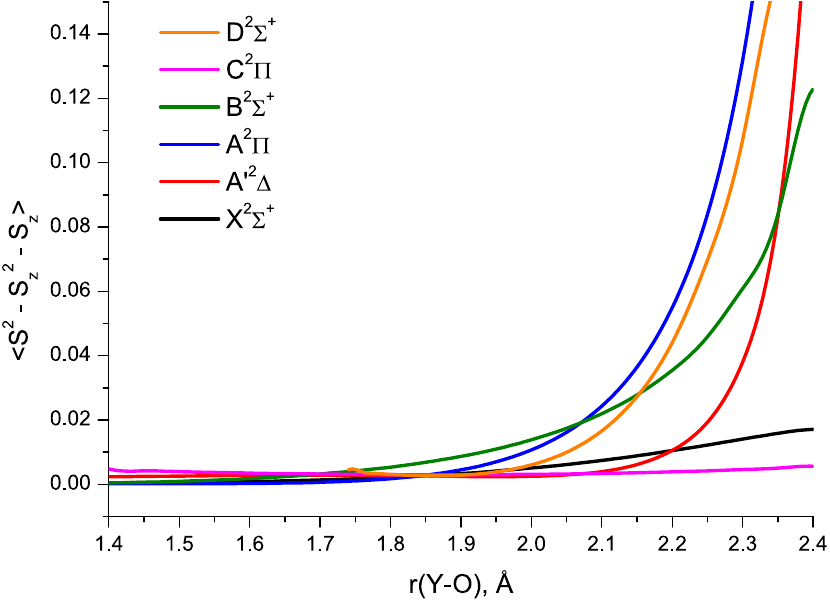}
	\caption{RHF-UCCSD/awCV5Z wave function spin contamination in the low-lying doublet electronic states of YO.}
  \label{f:SC}
\end{figure}

The CCSD/awCV5Z $T_1$ plots vs. Y--O bond length are shown in Fig.~\ref{f:T1}. The similar $D_1$ plots are
illustrated in Fig.~S1 of the supplementary material. At shorter bond lengths, the diagnostics amount to
0.02--0.03 ($T_1$) and 0.05--0.12 ($D_1$), remaining below the MR thresholds down to 1.4 \AA\ for most
states. Upon bond stretch, $T_1$ and $D_1$ rapidly increase, typically exceeding the MR threshold at 2.1--2.2
\AA. The behaviour of these diagnostics for the \C\ state is a notable exception: the numerical values of
both $T_1$ and $D_1$ remain well below the MR threshold throughout the bond length range studied. The \D\
state is also noteworthy: its $T_1$ and $D_1$ diagnostics are indicative of the CCSD \D\ wave function
retaining its SR character in much narrower range of bond lengths compared to the other doublet states under
study.

The relative importance of SR/MR character of YO can also be guessed with the use of spin contamination
appearing in RHF-UCCSD calculations as a result of unrestricted spin at the CCSD level. According to Jiang
{\it et al.},~\cite{12JiDeWi.ai} spin contamination with <$S^2$ $-$ $S_z^2$ $-$ $S_z$> greater than 0.1 in an
RHF-UCCSD wave function can be viewed as a strong indication of nondynamical correlation in an open-shell
system. Plotting spin contamination vs. bond length, Fig.~\ref{f:SC}, clearly indicates the severe admixture
of higher spin states in the CCSD \Ap, \A, \B\ and \D\ wave functions at bond lengths beyond 2.2--2.3 \AA.
Greater extent of spin contamination at longer internuclear distances can obviously be associated with larger
values of $T_1$ and $D_1$ (Fig.~\ref{f:T1} and Fig.~S1 of the supplementary material) exceeding the
established MR thresholds.

It is instructive to compare the MR diagnostics discussed above with the weights of the principal
configurations, $C{_0^2}$, in the MRCI wavefunctions of YO (see Table~\ref{t:wtsd}). At shorter bond lengths,
the $C{_0^2}$ values amount to $\sim$0.73 for the \B\ state and 0.79 -- 0.83 for the remaining doublet
electronic states under study. These values are smaller than the threshold, $C{_0^2}$ $\geq$ 0.90, proposed
in Refs.\citenum{12JiDeWi.ai,15WaMaWi.ai} to recognize the wave function strongly dominated by a single
configuration. It should, however, be noted that this criterion was established by analyzing the CASSCF
wavefunctions, whereas the $C{_0^2}$ of the entire MRCI wavefunction differs from that of the CASSCF
reference function due to the contributions of external configurations, which make $C{_0^2}$ a smaller
number.

Upon the YO bond stretch, there is a gradual decrease in the weights of the configurations serving as a
reference for the coupled cluster treatment of the \X, \Ap, and \A\ states. This indicates greater
multireference character of the respective wave functions at longer bond distances, as do the CC-based MR
diagnostics. The reference configuration for the \C\ state has approximately the same weight, $C{_0^2}$
$\cong$ 0.83, in the MRCI wavefunctions of YO throughout the bond length range explored, behaving just like
the respective CC-based MR diagnostics. These examples of the CCSD -- MRCI correlations imply that the
CC-based MR diagnostics can be capable of providing qualitative data about the relative accuracy in the
single-reference coupled cluster calculation results not only for near-equilibrium regions of electronic
ground states, but also for excited states in a wider range of molecular geometries.


In general, the present analysis indicates essentially single reference character of the YO low-lying doublet
states over most part of bond length range explored in our work and hence high accuracy in the respective
domains of the CCSD(T) PECs. It may also be indicative of accuracy degradation at larger bond lengths,
implying the need for additional adjustments of the CCSD(T) PECs. Nevertheless, the bond length range
associated with high-energy sections of PECs is expected to have a limited impact on the simulated spectra.

\subsubsection{Quartet states}

We have studied five low-lying quartet electronic states of YO at the CASSCF, CASPT2, CASPT3, and MRCI levels
of theory using the aVTZ basis set. The results are shown in Table~\ref{t:quart} together with the earlier
theoretical findings.\cite{88LaBaxx.YO} The lowest quartet, \aq, was also studied at the CCSD(T) level
(Table~\ref{t:prop_CC}). At larger internuclear distances, e.g., at $r = 2.19$~\AA, all the quartets feature
similar orbital character corresponding to the Y 5$s^1$4$d^1$, O 2p$^5$ electron configuration consistent
with the Y$^+$O$^-$ bonding (see Table~\ref{t:wtsq}).

The results for the YO quartet states obtained in our work agree with those of \citet{88LaBaxx.YO},
Table~\ref{t:quart}, except for the symmetry of the lowest quartet state that was reported\cite{88LaBaxx.YO}
to be $^{4}\Phi$ rather than $^{4}\Pi$.

The single-reference CCSD(T) method is expected to yield quite accurate results for the \aq\ state of YO
since its MR diagnostics, $C_0^2$ = 0.90, $T_1$ = 0.024, and $D_1$ = 0.078, indicate essentially SR character
of the \aq\ wave function in the vicinity of the \aq\ PEC minimum, 2.00 -- 2.25 \AA.
The very large CCSD(T)/CBS excitation energy of the \aq\ state, 29600 \cm, suggests that the quartet states
in YO are too high in energy to significantly affect the spectroscopy of its low-lying doublet states.

\begin{table}
\caption{Theoretical spin-free equilibrium constants of YO in its low-lying quartet states: adiabatic
excitation energy $T_\rme$ -- \cm, bond length $r_\rme$ -- \AA, vibrational frequency $\omega_\rme$ -- \cm. The aVTZ basis set has been used throughout.}
\label{t:quart} \centering
\begin{threeparttable}
\begin{tabular}{llrrrrr}
\hline
\hline
	& & $^{4}\Pi$ & $^{4}\Phi$ & $^{4}\Sigma^{+}$ & $^{4}\Delta$ & $^{4}\Sigma^{-}$ \\
\hline
$T_\rme$\tnote{a} & CASSCF         & 20460 & 68 & 1062 & 1825 & 2542 \\
& CASPT2    & 24140 & 56 & 1432 & 2158 & 2785 \\
& CASPT3    & 23772 & 53 & 1450 & 2175 & 2788 \\
& MRCI    & 25046   & 55 & 1394 & 2128 & 2748 \\
& MRCI+Q  & 26274   & 59 & 1357 & 2089 & 2699 \\
& CASSCF\cite{88LaBaxx.YO} & 26975\tnote{b,c} & 52 & 1933 & 2664 & 3359 \\
\hline
$r_\rme$ & CASSCF         & 2.141 & 2.143 & 2.110 & 2.110 & 2.114 \\
& CASPT2    & 2.197 & 2.198 & 2.192 & 2.191 & 2.196 \\
& CASPT3    & 2.210 & 2.211 & 2.207 & 2.208 & 2.214 \\
& MRCI    & 2.213 & 2.214 & 2.209 & 2.209 & 2.214 \\
& MRCI+Q  & 2.218 & 2.219 & 2.218 & 2.218 & 2.222 \\
& CASSCF\cite{88LaBaxx.YO} & 2.121\tnote{b} & 2.122 & 2.108 & 2.109 & 2.114 \\
& MCPF\cite{88LaBaxx.YO} & 2.126\tnote{b} &  & &  & \\
\hline
$\omega_\rme$ & CASSCF        & 517 & 516 & 522 & 521 & 519 \\
& CASPT2    & 515 & 515 & 517 & 494 & 477 \\
& CASPT3    & 519 & 518 & 524 & 494 & 474 \\
& MRCI    & 507 & 507 & 501 & 495 & 492 \\
& MRCI+Q  & 506 & 506 & 501 & 496 & 493 \\
& CASSCF\cite{88LaBaxx.YO} & 543 & 543 & 524 & 522 & 520 \\
& MCPF\cite{88LaBaxx.YO} & 526\tnote{b} & & &  & \\
\hline
\hline
\end{tabular}
\begin{tablenotes}
\small
\item[a] The energies of the $^{4}\Phi$, $^{4}\Sigma^{+}$, $^{4}\Delta$ and
    $^{4}\Sigma^{-}$ states are given here with respect to the $^{4}\Pi$ state which is the lowest-lying quartet state
    of YO.
\item[b] In Ref.\citenum{88LaBaxx.YO}, the symmetry of the lowest quartet state of YO was reported
      to be $^{4}\Phi$ rather than $^{4}\Pi$.
\item[c] In Ref.\citenum{88LaBaxx.YO}, the adiabatic excitation energy of the lowest quartet state was
    obtained from the MCPF calculations, and the relative energies of various quartet states were
    determined at the CASSCF level.
\end{tablenotes}
\end{threeparttable}
\end{table}

\begin{table*}
\caption{Main configurations in the low-lying quartet electronic states of YO derived from analyzing the MRCI/aVTZ
wave function at a bond length of 2.19 \AA. Weight of each configuration is $\sim$45$\%$.} \label{t:wtsq}
\centering
\begin{threeparttable}
\begin{tabular}{lcccccccccc}
\hline \hline
& \multicolumn{10}{c}{Configurations\tnote{a}} \\
State & 10$\sigma$ & 11$\sigma$ & 12$\sigma$ & 13$\sigma$ & 5$\pi_{+}$ & 5$\pi_{-}$ & 6$\pi_{+}$ & 6$\pi_{-}$ & 2$\delta_{+}$ & 2$\delta_{-}$ \\
\hline
$^{4}$$\Pi$ ($b_1$)       & 2 & 2 & + & 0 & + & 2 & 0 & 0 & + & 0 \\
                   & 2 & 2 & + & 0 & 2 & + & 0 & 0 & 0 & + \\
\hline
$^{4}$$\Phi$ ($b_1$)      & 2 & 2 & + & 0 & + & 2 & 0 & 0 & + & 0 \\
                   & 2 & 2 & + & 0 & 2 & + & 0 & 0 & 0 & + \\
\hline
$^{4}$$\Sigma^{+}$ & 2 & 2 & + & 0 & + & 2 & + & 0 & 0 & 0 \\
                   & 2 & 2 & + & 0 & 2 & + & 0 & + & 0 & 0 \\
\hline
$^{4}$$\Delta$ ($a_1$)    & 2 & 2 & + & 0 & + & 2 & + & 0 & 0 & 0 \\
                   & 2 & 2 & + & 0 & 2 & + & 0 & + & 0 & 0 \\
\hline
$^{4}$$\Sigma^{-}$ & 2 & 2 & + & 0 & + & 2 & 0 & + & 0 & 0 \\
                   & 2 & 2 & + & 0 & 2 & + & + & 0 & 0 & 0 \\
\hline \hline
\end{tabular}
\begin{tablenotes}
\small
\item[a] See footnote to Table~\ref{t:wtsd} for designations.
\end{tablenotes}
\end{threeparttable}
\end{table*}

\subsubsection{SO coupling}

Spin--orbit coupling effects were studied in a perturbative fashion at the MRCI level and more rigorously at
the 4c-EOM-CCSD level of theory including spin from the outset. The 4c-EOM-CCSD calculated spin--orbit
coupling effects on the spectroscopic constants of YO are shown in Table~\ref{t:scparam}. The theoretical
spin--orbit coupling constants, SOCCs ($A_{\rm SO}$), can be compared with the relevant experimental numbers for
the \Ap\ and \A\ electronic states of YO reported previously.\cite{76ChGoxx.YO,83BeGrxx.YO,07BaCaG1.YO} The
$A_{\rm SO}$(\Ap) SOCCs of 336 \cm\ and 313 \cm\ obtained at the MRCI SI-SO and 4c-EOM-CCSD levels, respectively,
agree well with each other and with the experimental number of 339 \cm\ determined by \citet{76ChGoxx.YO}.
The calculation results for $A_{\rm SO}$(\A), 346 \cm\ (MRCI SI-SO) and 438 \cm\ (4c-EOM-CCSD), are also in
reasonable agreement with the respective experimental value of 428 \cm.\cite{83BeGrxx.YO,07BaCaG1.YO} As the
Y--O distance reaches the avoided crossing point between \A\ and \C, the $A_{\rm SO}$(\C) and $A_{\rm SO}$(\A) values
change their sign: the $A^{2}\Pi_{3/2}$ spin component of the \A\ state becomes lower in energy than its
$A^{2}\Pi_{1/2}$ counterpart, and vice versa for the spin--coupled components of the \C\ state (see
Fig.~\ref{f:eom-so}). There is also a change in the absolute values of $A_{\rm SO}$(\A) and $A_{\rm SO}$(\C): at $r$
< $r_{ac}$, |$A_{\rm SO}$(\A)| is much lower in magnitude than |$A_{\rm SO}$(\C)| and vice versa at $r$ > $r_{ac}$.
However, the numerical values of $A_{\rm SO}$(\A) at $r$ > $r_{ac}$ determined with the MRCI SI-SO and
4c-EOM-CCSD methods, e.g., $-$45 \cm\ and $-$186 \cm, respectively, at a bond length of 2.04 \AA, are in less
satisfactory agreement with each other than those at $r$ < $r_{ac}$.

\begin{figure}
  \includegraphics[width=0.46\textwidth]{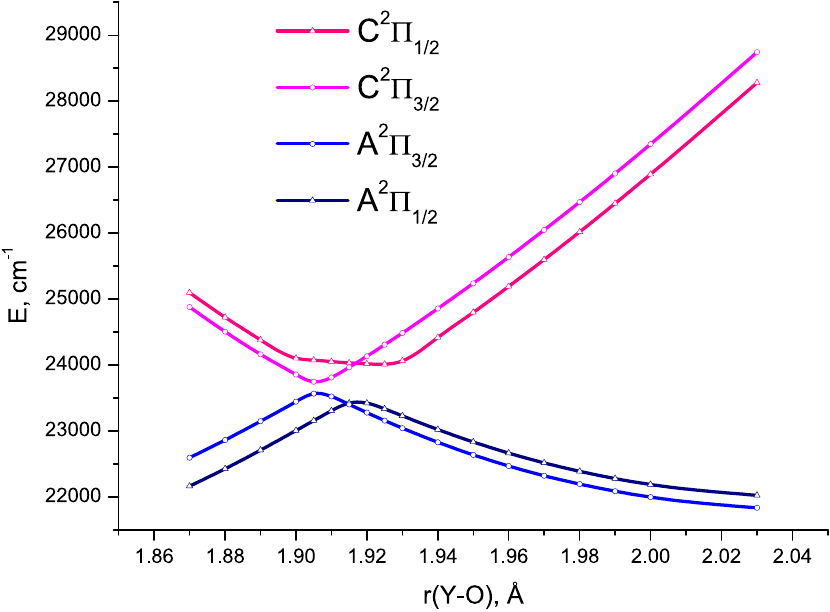}
	\caption{4c-EOM-IP-CCSD/TZ$_D$ potential energy curves for the spin--coupled components of the \A\ and \C\ electronic
states of YO in the avoided crossing region of bond length values.}
  \label{f:eom-so}
\end{figure}

The SOC matrix elements between various doublet states of YO, which also accurately account for the
corresponding phases, are shown in Fig.~\ref{f:socs_new} as a function of $r$(Y--O). The relative phases of
the couplings are important when used for solving the nuclear motion problem as part of the coupled
Schr\"{o}dinger equation, see, for example, discussion by \citet{jt589} The full details of the \ai\ coupling
curves including the magnetic quantum numbers are provided as part of the supplementary data.

\subsubsection{Dipole moment, transition dipole moment, and electronic angular momentum curves of YO}

The CCSD(T)/CBS dipole moment of 4.61~D for the \X\ state of YO (Table \ref{t:prop_CC}) is in agreement with
the respective values obtained experimentally by \citet{90StShxx.YO} in a molecular beam-optical Stark study,
4.45(7)~D, and by \citet{90SuLoFr.YO} from the more precise microwave measurement, 4.524(7)~D. For the
spin--orbit components $\Omega$ = 1/2 and $\Omega$ = 3/2 of the \A\ state, the dipole moment values,
$\mu_e$($A^{2}\Pi_{1/2}$) = 4.185~D and $\mu_e$($A^{2}\Pi_{3/2}$) = 4.125~D, were obtained at the
4c-EOM-EA-CCSD/TZ$_D$ level of theory at the respective CCSD(T)/CBS equilibrium bond lengths of 1.7937 \AA\
and 1.7929 \AA, estimated by applying the 4c-EOM-EA-CCSD $\Delta_{\rm SO}$$r_\rme$ spin--orbit corrections (from
Table~\ref{t:scparam}) to the spin-free CCSD(T)/CBS1 bond length, $r_\rme$ = 1.7932 \AA. The 4c-EOM-EA-CCSD
dipole moments are overestimated by 0.4--0.5 D compared to the spin-free CCSD(T)/CBS $\mu_e$(\A) value of
3.73~D, the latter being in good agreement with the $\mu_e$($A^{2}\Pi_{3/2}$) = 3.68(2)~D measured by
\citet{90StShxx.YO}. However, the experimental work\cite{90StShxx.YO} reports the dipole moment
$\mu_e$($A^{2}\Pi_{1/2}$), 3.22(8)~D, to be lower than $\mu_e$($A^{2}\Pi_{3/2}$). This result is not
supported by our \ai\ calculations. \citet{90StShxx.YO} compared the dipole moments in the \A\ spin--orbit
components of YO to those of the valence-isoelectronic molecule ScO\cite{90ShScSt.ScO}, where
$\mu_e$($A^{2}\Pi_{1/2}$) > $\mu_e$($A^{2}\Pi_{3/2}$), and proposed an explanation of the different order in
YO. According to \citet{90StShxx.YO}, the reason for $\mu$($A^{2}\Pi_{3/2}$) being larger than
$\mu$($A^{2}\Pi_{1/2}$) in YO in contrast to ScO is the smaller energy gap of the \A\ and \Ap\ states, which
results in mixing between the $\Omega$ = 3/2 spin--orbit components of these states. This idea is, however,
based on low-level ab initio computations by \citet{88LaBaxx.YO}, which predicted the \Ap\ state in YO to lie
~200 \cm\ higher than \A, whereas for ScO the analogous calculations\cite{86BaLaxx.ScO} yielded a difference
of about 1900 \cm, with \Ap\ being lower in energy. In fact, the \Ap\ state  lies around 1800 \cm\ lower than
\A\ in YO and 1500 \cm\ lower in ScO, as evidenced by experimental works of Chalek and
Gole\cite{76ChGoxx.YO,77ChGoxx.YO}, i.e., the \A\ -- \Ap\ energy gap in YO exceeds that in ScO. Also, at high
levels of theory including SO coupling, the PECs for the $\Omega$ = 3/2 components of \A\ and \Ap\ lie quite
far apart (see the excitation energies in Table~\ref{t:scparam}), and their mixing is almost negligible.
Furthermore, it is worth noting that for another analogous molecule, LaO, the experimental
data\cite{02StVixx.LaO} also indicate that $\mu_e$($A^{2}\Pi_{1/2}$) > $\mu_e$($A^{2}\Pi_{3/2}$).

\begin{figure}
  \includegraphics[width=0.42\textwidth]{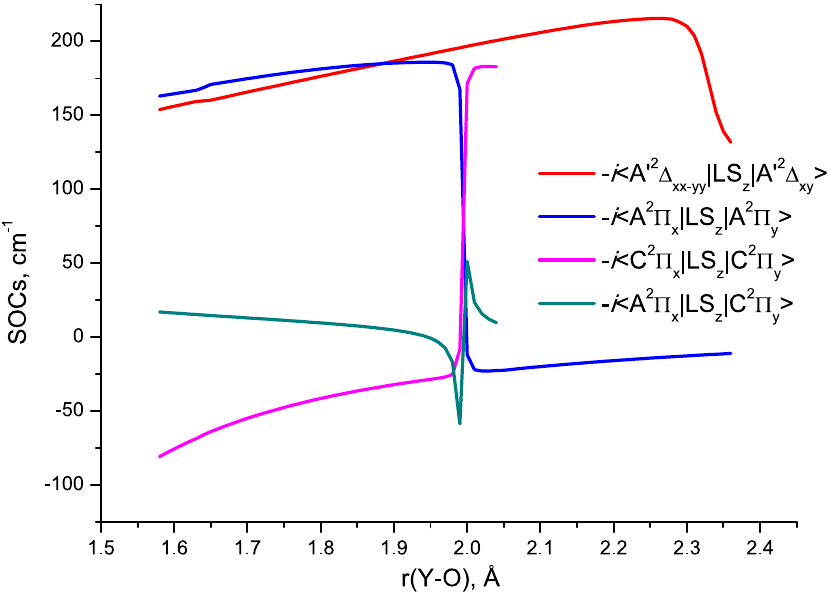}
  \includegraphics[width=0.42\textwidth]{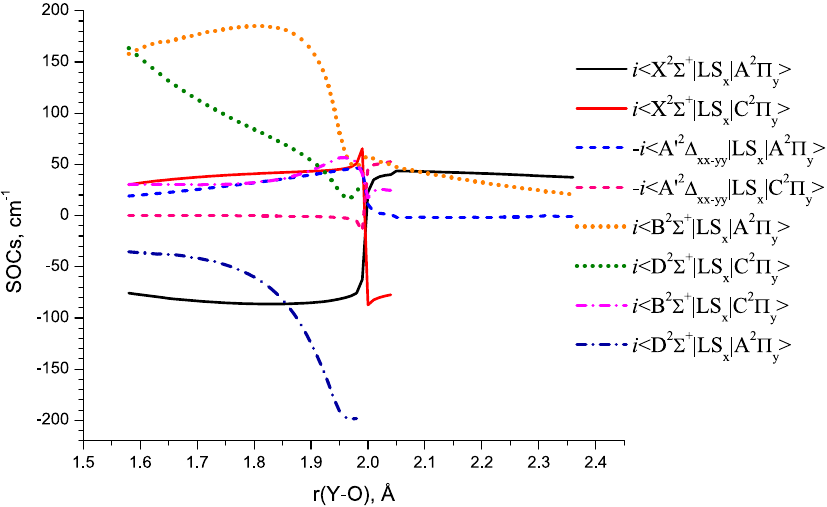}
	\caption{The MRCI/aVTZ SI-SO calculated <i|LS$_z$|j> (above) and <i|LS$_x$|j> (below) spin--orbit matrix elements of YO.}
  \label{f:socs_new}
\end{figure}

To shed more light on the alleged different order of the dipole moment values for the spin--orbit components
of the \A\ state in YO compared to ScO and LaO, we performed additional 4c-EOM-EA-CCSD/TZ$_D$ computations
for the two latter molecules at the experimental equilibrium bond lengths\cite{72StAtFe.ScO,00BeVexx.LaO} of
1.6826 \AA\ (ScO) and 1.8400 \AA\ (LaO). These resulted in the following values: $\mu$($A^{2}\Pi_{1/2}$) =
4.543~D, $\mu$($A^{2}\Pi_{3/2}$) = 4.532~D (ScO), $\mu$($A^{2}\Pi_{1/2}$) = 3.011~D, $\mu$($A^{2}\Pi_{3/2}$)
= 2.907~D (LaO), i.e., the \ai\ predicted difference between the two spin--orbit components monotonically
increases on passing in the series ScO $\rightarrow$ YO $\rightarrow$ LaO: 0.01 $\rightarrow$ 0.04
$\rightarrow$ 0.10 D, respectively. The experimental counterparts\cite{90ShScSt.ScO,02StVixx.LaO} are:
$\mu$($A^{2}\Pi_{1/2}$) = 4.43(2)~D, $\mu$($A^{2}\Pi_{3/2}$) = 4.06(3)~D (ScO), $\mu$($A^{2}\Pi_{1/2}$) =
2.44(2)~D, $\mu$($A^{2}\Pi_{3/2}$) = 1.88(6)~D (LaO). Since the 4c-EOM-EA-CCSD dipole moments are expected to
be overestimated by at least 0.5 D, one can consider the theoretical results to be in reasonable agreement
with experiment. In light of our results, the experimental dipole moments\cite{90StShxx.YO} for the two
$\Omega$ components of the \A\ state of YO need to be revisited.

The MCPF dipole moments obtained by \citet{88LaBaxx.YO}, 3.976~D (\X), 7.493~D (\Ap) and 3.244~D (\A), are
systematically smaller than our CCSD(T) (Table~\ref{t:prop_CC}), MRCI, and EOM-EA-CCSD (Table~\ref{t:sfree})
results.

The MRCI DMCs and TDMCs of YO are shown in Fig.~\ref{f:dmcs_new}. The EAMCs are shown in
Fig.~S2 of the supplementary material. All these curves as well as the SOC ones (Fig.~\ref{f:socs_new})
exhibit irregular behaviour at bond lengths around $r\sim$2~\AA\ due to strong changes in the \A, \B, \C\ and
\D\ wave functions over the avoided crossing region.

\begin{table*}
\caption{4c-EOM-CCSD/TZ$_D$ molecular properties of YO in its low-lying spin-coupled electronic states (EOM-IP for
\C\ and EOM-EA for the remaining states):
 bond length $r_\rme$ -- \AA, vibrational frequency $\omega_\rme$,
 adiabatic excitation energy $T_\rme$ -- \cm. The respective spin--orbit effects,  $\Delta_{\rm SO}$,
 are provided as well.}
\label{t:scparam}
\centering
\begin{tabular}{lcrcrrr}
\hline
\hline
& $r_\rme$ & $\Delta_{\rm SO}$$r_\rme$\hspace{5pt} & $\omega_\rme$ & $\Delta_{\rm SO}$$\omega_\rme$
& $T_\rme$\hspace{10pt} & $\Delta_{\rm SO}$$T_\rme$\hspace{5pt} \\
$X^{2}\Sigma^{+}_{1/2}$ & 1.7866 &   0.0000 & 880.9 & 0.0   & 0     & 0 \\
$A'^{2}\Delta_{3/2}$    & 1.8109 & +0.0003  & 821.5 & $-$0.5 & 15937 & $-$159 \\
$A'^{2}\Delta_{5/2}$    & 1.8102 & $-$0.0004 & 822.6 & +0.6  & 16254 & +157 \\
$A^{2}\Pi_{1/2}$        & 1.7892 & +0.0005	& 846.9 & $-$0.5 & 16591 & $-$226 \\
$A^{2}\Pi_{3/2}$        & 1.7884 & $-$0.0003 & 847.8 & +0.4  & 17028 & +212 \\
$B^{2}\Sigma^{+}_{1/2}$ & 1.8181 & $-$0.0003 & 804.5 & +0.6  & 21671 & +17 \\
$C^{2}\Pi_{3/2}$        & 2.0488 & $-$0.0004 & 605.7 & +0.1  & 21808 & $-$88 \\
$C^{2}\Pi_{1/2}$        & 2.0495 &  +0.0003	& 605.6 & $-$0.0 & 21994 & +98 \\
\hline
\hline
\end{tabular}
\end{table*}

\subsection{Results of \Duo\ calculations}

For \Duo\ calculations, we selected the following set of curves representing our highest level of theory: the
CCSD(T)/CBS PECs shown in Fig.~\ref{f:PECsCC} and MRCI SOCs, (T)DMCs, and EAMCs shown in
Figs.~\ref{f:socs_new} and \ref{f:dmcs_new}, as well as Fig.~S2 of the supplementary material. Due to
limitations of single reference CCSD(T), the CCSD(T) curves for the \A, \C\ and \B, \D\ states do not exhibit
avoided crossings and hence are not consistent with the MRCI property curves. To alleviate this deficiency,
these four PECs were transformed by simply switching the corresponding energy points between $A$ and $C$
($^2\Pi$ states) as well as those between $B$ and $D$ ($^2\Sigma^+$ states) at $r$
> $r_{ac}$. We have decided to apply this rather simplistic procedure because it has marginal effect on
the overall accuracy of our model and is sufficient for the goal of this pure \ai\ study, not aiming at
spectroscopic accuracy. A proper diabatic representation of the YO electronic states will be, however,
important when refining the \ai\ curves by fitting to experiment, \cite{jt736} which is a goal of future
work. In this study we work directly with the \ai\ data in the grid representation without representing \ai\
curves analytically. We do not perform any diabatizations here, which is often useful for representing the
variation of the data with respect to the bond length in an intuitive and more compact form. One of the
artifacts of this choice to use the \ai\ curves directly is that the crossing points of the MRCI PECs hence
the points of drastic change in the MRCI property curves differ by a few hundredths of \AA\ from the crossing
points of the CCSD(T)/CBS PECs. Again, this has small impact on the overall agreement of the current model
with the experiment. However, a more accurate study will require a more consistent treatment of the crossing
points. Our preferred choice would be to use the CCSD(T)/CBS values of the corresponding crossing points.

\begin{figure}[H]
	\includegraphics[width=0.40\textwidth]{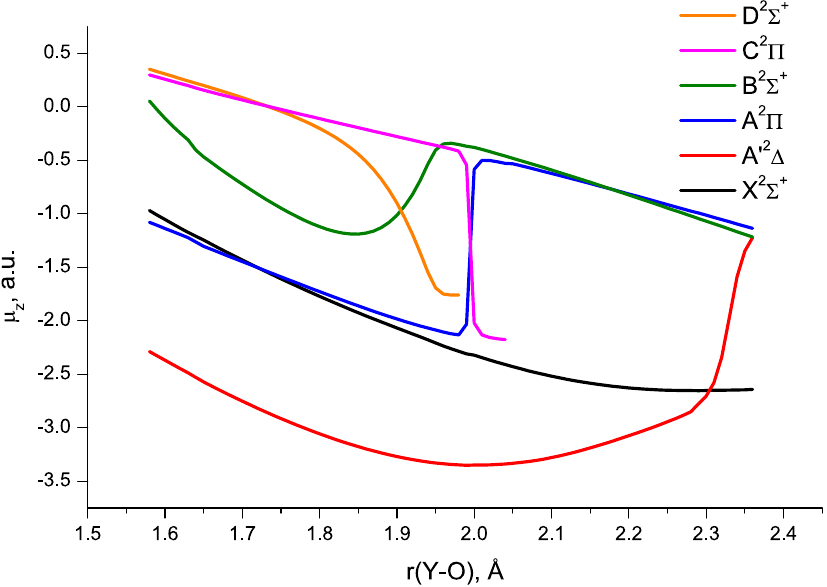}
	\includegraphics[width=0.40\textwidth]{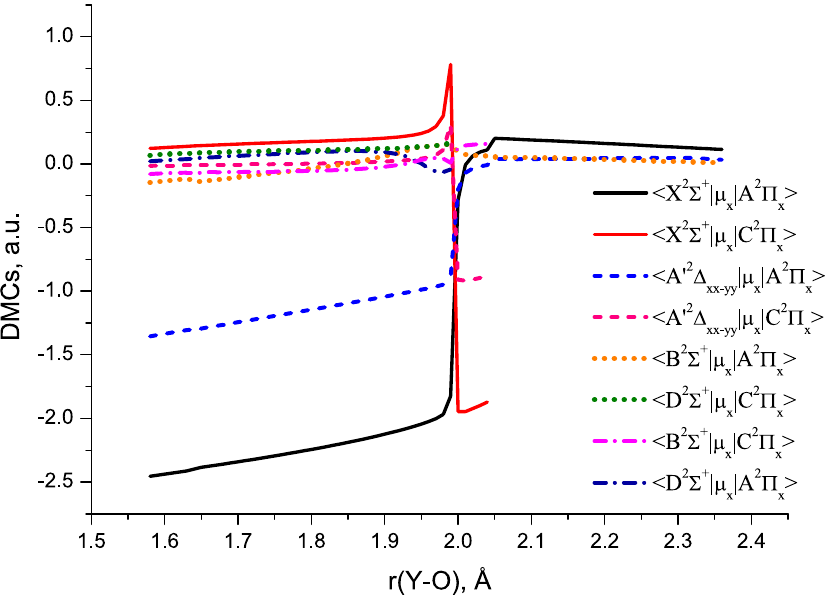}
	\includegraphics[width=0.40\textwidth]{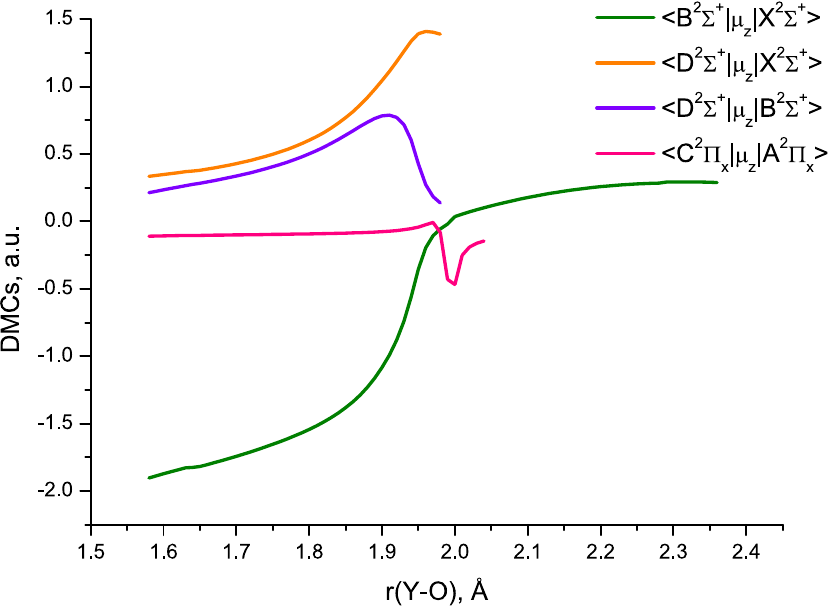}
	\caption{MRCI/aVTZ electric dipole moment curves of YO: diagonal (above), non-diagonal $\mu_x$ (middle) and  non-diagonal  $\mu_z$ (below). }
	\label{f:dmcs_new}
\end{figure}

The \Duo\ rovibronic wavefunctions of YO in conjunction with the \ai\ TDMCs were then used to produce
Einstein~$A$ coefficients for all rovibronic transitions between states considered in this work covering the
wavenumber range from 0 to 40000~\cm\ and $J \le 180.5$. These Einstein~$A$ coefficients and the
corresponding energies from the lower and upper states involved in each transition were organized as a line
list using the ExoMol format.\citep{jt631} This format uses a two file structure with the energies included
into the States file (.states) and Einstein coefficients appearing in the Transitions file (.trans). This
\ai\ line list is available from \verb!www.exomol.com!. The ExoMol format has the advantage of being compact and
compatible with our intensity simulation program \textsc{ExoCross} \cite{jt708} (see below).

\subsubsection{Vibronic energies}

In Table~\ref{t:band-centers} we compare our computed vibrational excitations at $J=0.5$ and $J=1.5$ (as
proxy for vibrational band centres) of $^{89}$Y$^{16}$O with the experimentally derived values. Based on this
comparison, as an ad hoc improvement we applied the following shifts to PECs of the excited states: +9.509
\cm\ (\A), +81.096 \cm\ (\Ap), $-$134.301 \cm\ (\B), and +358.626 \cm\ (\D). These shifts
are small compared to the observed minus calculated differences often encountered in calculations of
electronic term values for transition metal oxides.\cite{jt632} We also scaled the SOC of \A\ by 1.14 in
order to increase the SO splitting of $v=0$ by about 33 \cm.  Even though we are not targeting fully
quantitative accuracy in this work, without such empirical shifts it would be difficult to reproduce band
heads even qualitatively.

\begin{table}
  \caption{Comparison of our \ai\ and experimentally derived term values of $^{89}$Y$^{16}$O in \cm. The \ai\
  PECs were shifted by +9.509 \cm\ (\A), +81.096 \cm\ (\Ap), $-$134.301 \cm\ (\B) and +358.626 \cm\ (\D).
  The SOC of \A\ was scaled by 1.1376. The `Obs' values of $A$, $A'$, $B$ and $D$ were derived using
  spectroscopic constants from the corresponding works with the help of the \textsc{PGOPHER} program\protect\cite{PGOPHER}. The $X$ state `Obs.' values are represented by the corresponding band centers (limit $J=0$).}
\label{t:band-centers}
  \centering
  \begin{tabular}{ccccrrr}
    \hline
    \hline
    &  $ \upsilon $      &   $J$    &    $\Omega$     &   \Duo & \multicolumn{2}{c}{Obs.} \\
    \hline
     &     &       &       &      &[\onlinecite{98ReNaRa.YO}] & [\onlinecite{83BeGrxx.YO}]\\
$X^a$  &   0 &       0 &          &              0 &                           0 &                      0   \\
     &   1 &       0 &          &        860.879 &                       855.2 &          855.7463(52)    \\
     &   2 &       0 &          &       1716.156 &                      1704.4 &         1705.8339(90)    \\
     &   3 &       0 &          &       2565.836 &                      2547.9 &         2550.2684(65)    \\
     &   4 &       0 &          &       3409.931 &                      3385.4 &         3389.0242(90)    \\
     &   5 &       0 &          &       4248.450 &                      4217.1 &          4222.085(11)    \\
     &   6 &       0 &          &       5081.364 &                             &          5049.454(13)    \\
$A$  &     &         &          &                &  [\onlinecite{83BeGrxx.YO}] &                          \\
     &   0 &     0.5 &   0.5    &      16295.492 &          16295.453          &                          \\
     &   0 &     1.5 &   1.5    &      16724.499 &          16724.541          &                          \\
     &   1 &     1.5 &   1.5    &      17117.400 &          17109.845          &                          \\
     &   1 &     1.5 &   1.5    &      17545.141 &          17538.459          &                          \\
     &   2 &     1.5 &   1.5    &      17931.536 &          17916.880          &                          \\
     &   2 &     1.5 &   1.5    &      18358.848 &          18345.768          &                          \\
     &   3 &     1.5 &   1.5    &      18740.505 &          18716.674          &                          \\
     &   3 &     1.5 &   1.5    &      19169.245 &          19146.593          &                          \\
     &   4 &     1.5 &   1.5    &      19547.495 &          19510.064          &                          \\
     &   4 &     1.5 &   1.5    &      19964.360 &          19940.488          &                          \\
     &   5 &     1.5 &   1.5    &      20350.928 &          20296.636          &                          \\
     &   5 &     1.5 &   1.5    &      20732.561 &          20727.595          &                          \\
     &     &       &         &                 &  [\onlinecite{92SiJaHa.YO}]   &                            \\
$B$  &   0 &     2.5 &   1.5    &      14500.074 &                   14502.010 &                          \\
     &     &       &         &                 &   [\onlinecite{79BeBaLu.YO}]  &  [\onlinecite{80BeGrxx.YO}] \\
     &   0 &     0.5 &   0.5    &      20741.630 &                   20741.688 &             20741.6877   \\
     &   1 &     0.5 &   0.5    &      21516.945 &                             &             21492.4773   \\
     &   2 &     0.5 &   0.5    &      22294.044 &                             &                          \\
     &   3 &     0.5 &   0.5    &      23102.252 &                             &               22941.71   \\
     &   4 &     0.5 &   0.5    &      23850.742 &                             &                23615.3   \\
$D$  &     &         &          &                &  [\onlinecite{17ZhZhZh}]    &                          \\
     &   0 &     0.5 &   0.5    &      23969.916 &                   23969.940 &                          \\
     &   1 &     0.5 &   0.5    &      24659.745 &                   24723.766 &                          \\
     \hline\hline
  \end{tabular}

  $^a$ Band centers.

\end{table}

To allow for a direct comparison with the observed spectra of $^{89}$Y$^{16}$O, we generated a line list covering
rotational excitations up to $J=190$ and the energy/wavenumber range up to 40,000~\cm, with a lower state
energy cutoff of 16,000~\cm.

\subsection{Partition function}

The partition function of $^{89}$Y$^{16}$O computed using our \ai\ line list is shown in Fig.~\ref{f:pf},
which is compared to that recently reported by \citet{16BaCoxx.partfunc}. Since $^{89}$Y has a nuclear spin
degeneracy of two, we have multiplied Barklem and Collet's partition function by a factor of two to
compensate for the different conventions used; we follow HITRAN\cite{jt692} and include the full nuclear spin
in our partition functions. The partition function of YO was also reported by \citet{70Vardya.YO}, which is
shown in Fig.~\ref{f:pf}. All three partition functions are almost identical for their ranges of validity.

\subsection{Spectral comparisons}

Using the \ai\ $^{89}$Y$^{16}$O line list, spectral simulations were performed with our code
\textsc{ExoCross}.\cite{jt708} \textsc{ExoCross} is an open source Fortran 2003 code with the primary use to
produce spectra of molecules at different temperatures and pressures in the form of cross sections using
molecular line lists as input. Here we use the YO line list generated with \Duo\ in the ExoMol format, the
description of which can be found, e.g., in  \citet{jt708} or \citet{jt631}. \textsc{ExoCross} can be
accessed via \verb!http://exomol.com/software/! or directly at \verb!https://github.com/exomol!. Amongst
other features, \textsc{ExoCross} can generate spectra for non-local thermal equilibrium conditions
characterized with different vibrational and rotational temperatures, lifetimes, Land\'e $g$-factors,
partition and cooling functions.

\begin{figure}
	\includegraphics[width=0.5\textwidth]{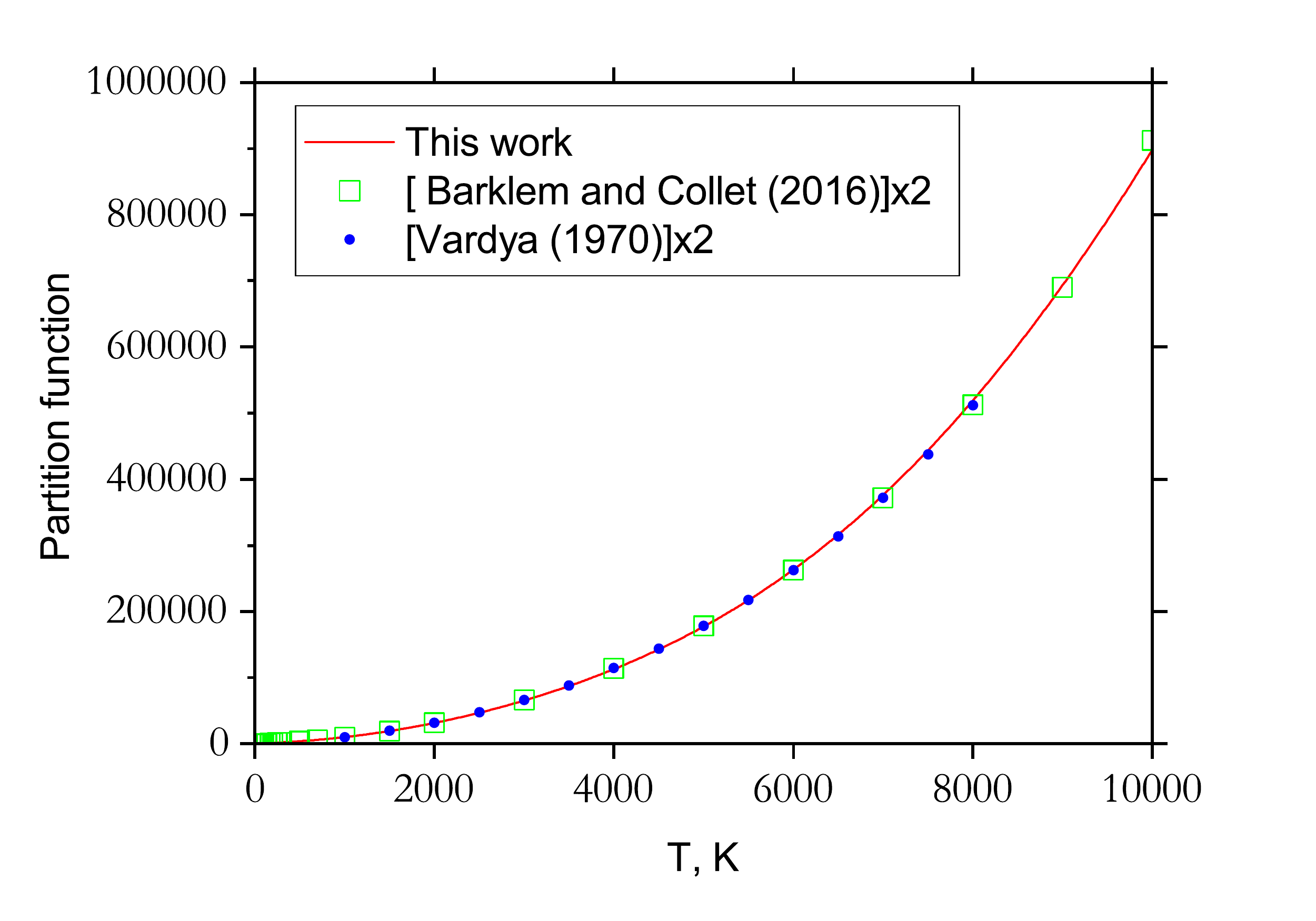}
	\caption{Partition functions of YO: Solid line is from this work computed using the energies of
  the six lowest electronic states;  filled circles represent the partition function values by
  \protect\citet{70Vardya.YO} generated using spectroscopic constants of 3 lowest electronic
  states $X$, $A$ and $B$ (multiplied by a factor of 2 to account for the nuclear statistics);
  open squares represent values by \protect\citet{16BaCoxx.partfunc} (times the factor 2). }
	\label{f:pf}
\end{figure}

An overview of the YO absorption spectra in the form of cross sections at the temperature $T=$ 2000~K is
illustrated in Fig.~\ref{f:T=2000K}. Here, a Gaussian line profile with a half-width-at-half-maximum (HWHM)
of 5~\cm\ was used.  This figure shows contributions from each electronic band originating from the ground
electronic state. The strongest bands are \A--\X\ and \B--\X. The visible $A$--$X$ band is known to be
important for the spectroscopy of cool stars. The $C$ state is of the same symmetry as $A$, however, the
corresponding band $C$--$X$ is much weaker due to the small Franck-Condon effects. The \Ap--\X\ band is
forbidden and barely seen in Fig.~\ref{f:T=2000K}, however, it is strong enough to be experimentally
known.\cite{92SiJaHa.YO}

Figure~\ref{f:T=3000K:A} shows a simulated emission spectrum of the strongest orange system YO (\A--\X,
(0,0)), which is compared to the experiment of \citet{02BaGrxx.YO} (from the plume emission close to the
liquid Y$_2$O$_3$ surface).  It is remarkable that even pure \ai\ calculations (after modest adjustment of
the corresponding $T_{\rm e}$ value by +9.509 \cm) provide very close reproduction of experiment. It shows
that our line list at the current, \ai\ quality should be useful for modelling spectroscopy of exoplanets and
cool stars in the visible region.

\begin{figure}[H]
	\includegraphics[width=0.41\textwidth]{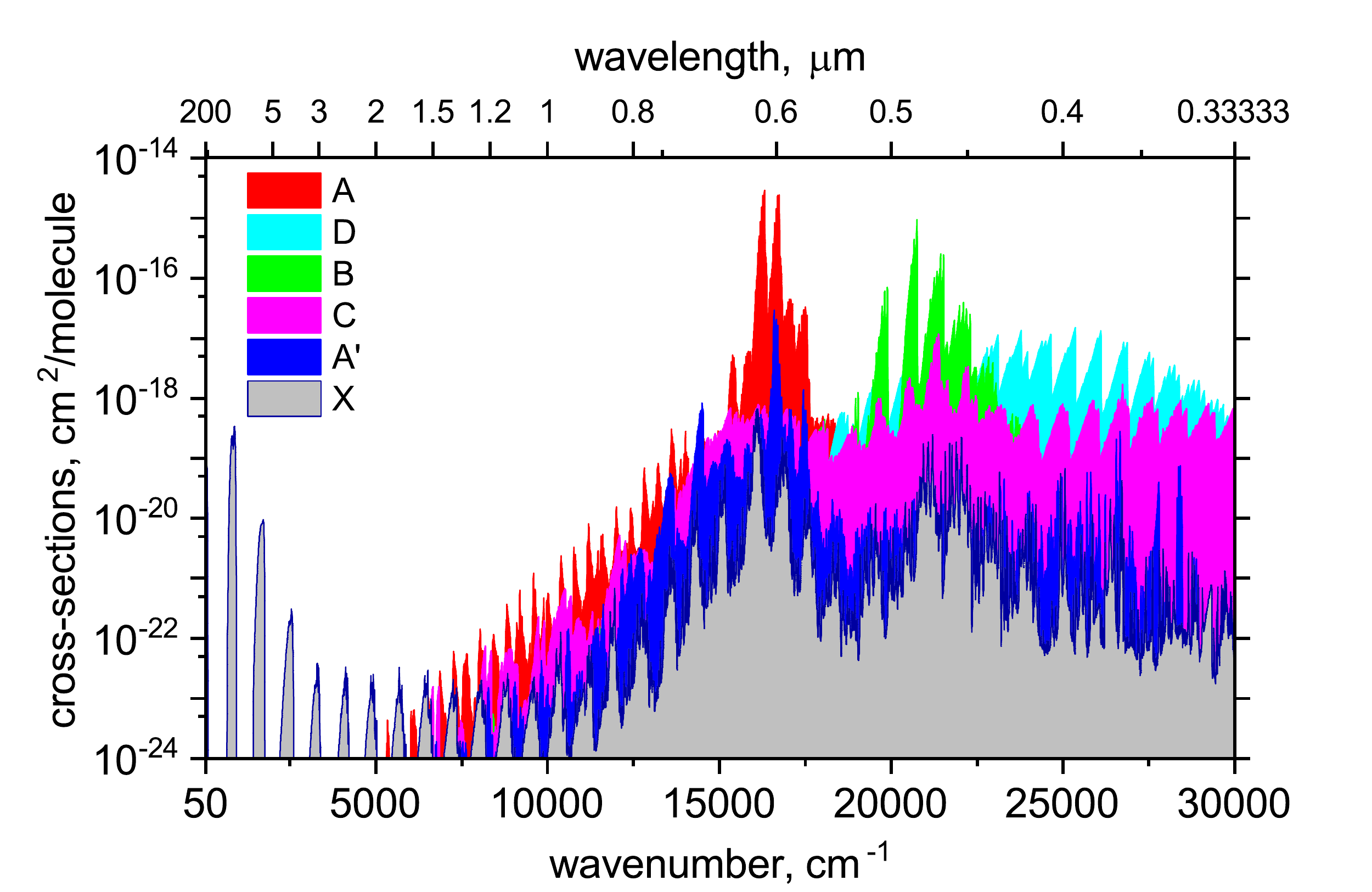}
	\caption{An overview of a theoretical absorption spectrum of YO at $T=2000$~K for different electronic bands, designated by their upper state.
	The spectrum was computed using our \ai\ line list for YO assuming a Gaussian profile with a half-width-at-half-maximum (HWHM) of 5~\cm.}
	\label{f:T=2000K}
\end{figure}

Figure \ref{f:T=77K:Ap} illustrates the \Ap\ -- \X\ (0,0) forbidden band in emission simulated for $T=77$~K
compared to the experimental spectrum of  \citet{92SiJaHa.YO}. Here, a shift of +81.096 \cm\ was applied to
the $T_{\rm e}$ value of the \Ap\ state. In spectral simulations, this region appears to be contaminated by
the dipole-allowed hot $A$--$X$ transitions, which are not necessarily very accurate in this region. We
therefore applied a filter to select the \Ap\ -- \X\ transitions only. The difference in shape of the spectra
can be attributed either to the non-LTE (Local Thermal Equilibrium) effects present in the experiment or
broadening effects, which we have not attempted to model properly. This figure is only to illustrate the
generally good agreement of the positions of the rovibronic lines in this band.

\begin{figure}
	\includegraphics[width=0.41\textwidth]{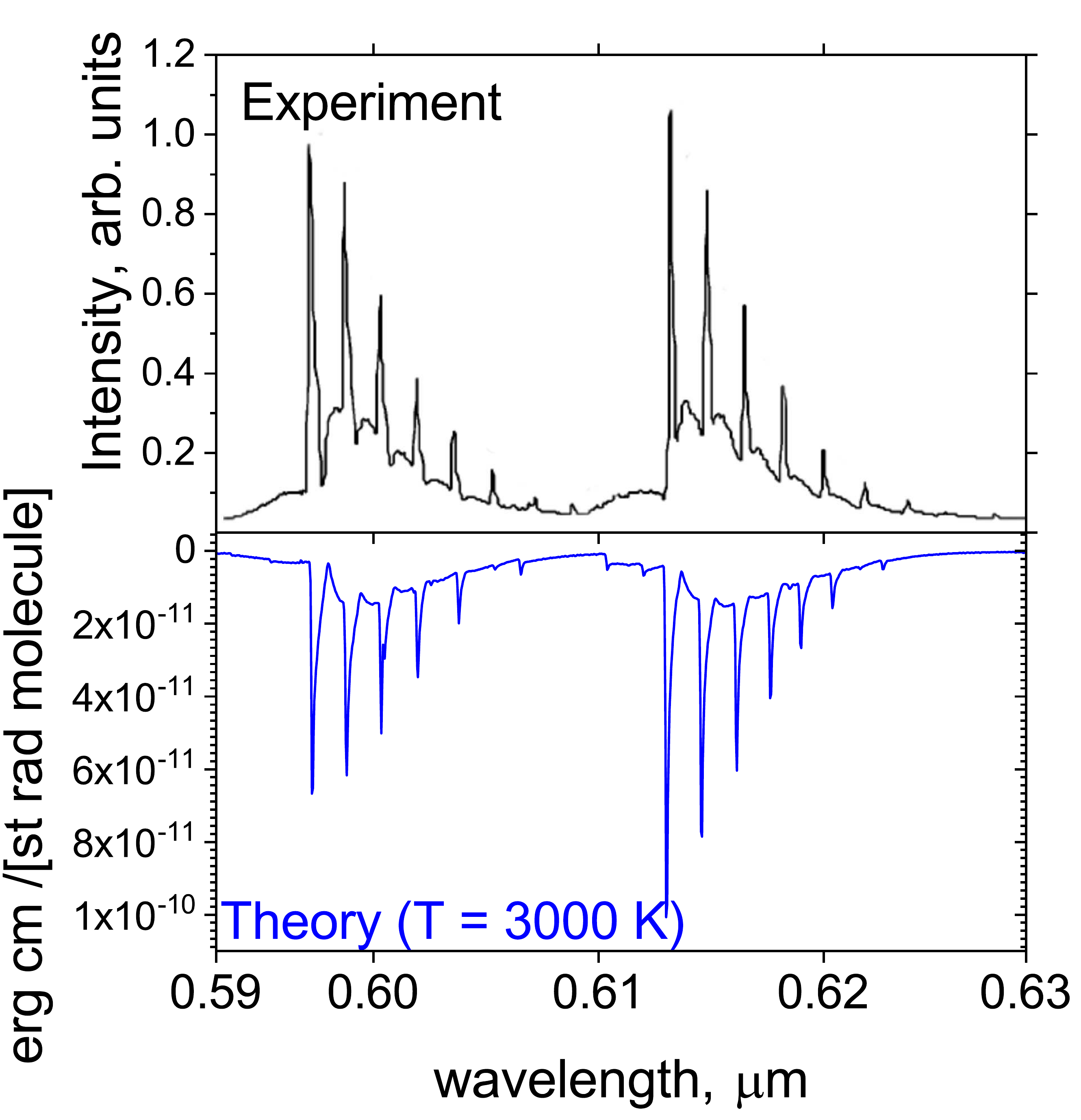}
	\caption{Comparison of the computed  \A\ -- \X\ orange band with the observations of \citet{02BaGrxx.YO}. Our simulations assume $T=3000$~K and Gaussian line profile of HWHM = 1~\cm.}
	\label{f:T=3000K:A}
\end{figure}

\begin{figure}[H]
	\includegraphics[width=0.5\textwidth]{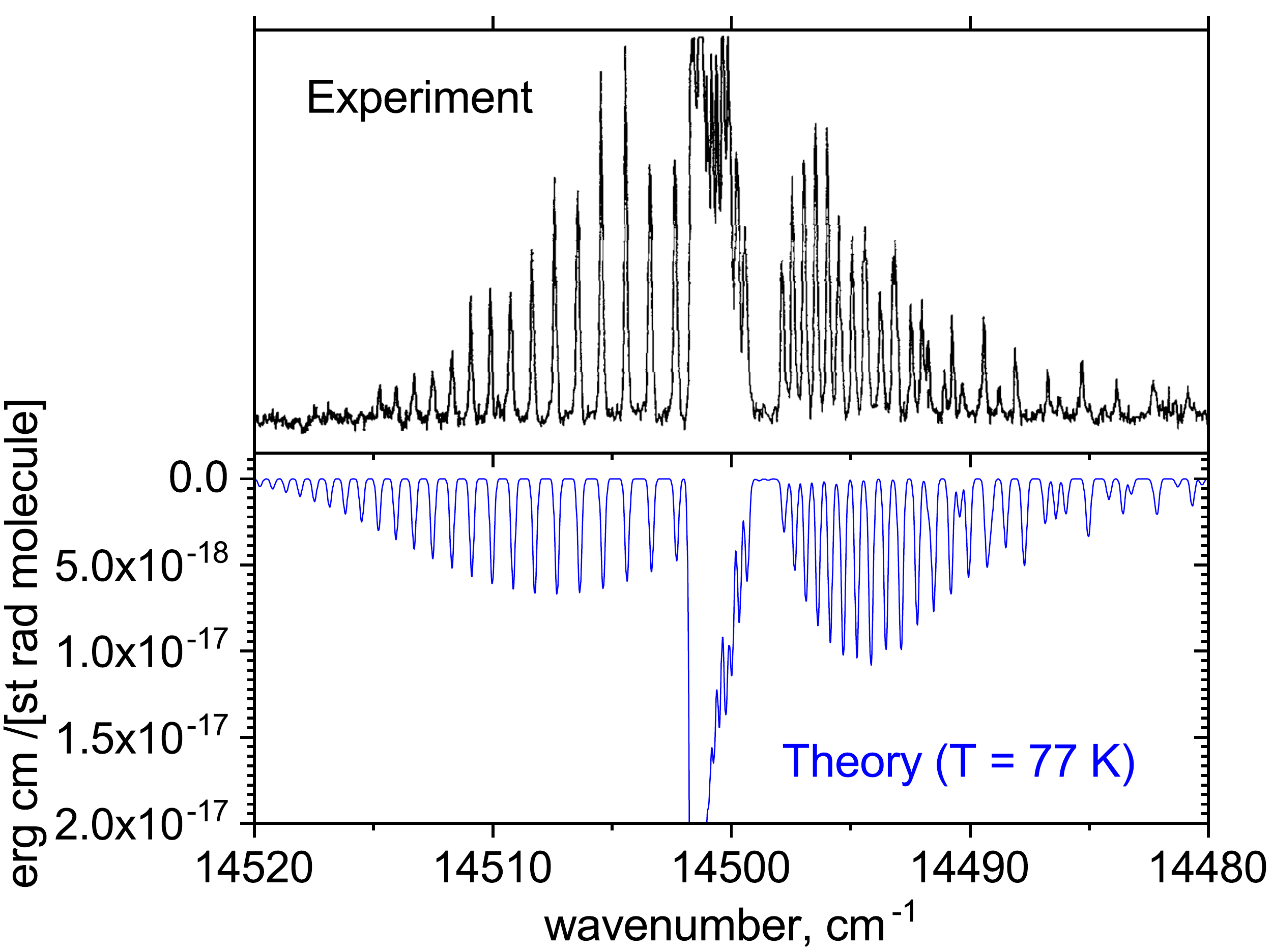}
	\caption{Comparison of the computed emission \Ap\ -- \X\ (0,0) band with the measurements of
\protect\citet{92SiJaHa.YO} at $T=77$ K
and Gaussian line profile of HWHM = 0.1~\cm.}
	\label{f:T=77K:Ap}
\end{figure}

Figure \ref{f:T=77K:2017} shows a series of absorption bands compared to the measurements of \citet{17ZhZhZh}
who observed bands in both the \B\ -- \X\ and \D\ -- \X\ systems in a heavily non-thermal environment where
the vibrations were hot and the rotations cooled to liquid nitrogen temperatures. In this case of multi-band
system it was important to include at least some non-LTE effects by treating it using two temperatures,
vibrational and rotational, assuming that the corresponding degrees of freedom are in LTE. The rotational
temperature $T_{\rm rot} = 77$~K was set to value specified by \citet{17ZhZhZh}, while the vibrational
temperature was adjusted to $T_{\rm vib}$  = 2000~K to better reproduce the experimental spectrum. The
spectrum is divided into five spectroscopic windows (I--V) which are also detailed in
Table~\ref{t:five:bands}. In order to match the positions of the vibronic bands in the experiment, some of
the windows were shifted. For example, the \D\ -- \X\ (1,0) band was shifted by about 76.5~\cm. This shift is
an indication of the inaccuracy of our model to reproduce the vibrationally excited states of \D. This is not
surprising considering the complexity of the quantum-chemistry part of these systems as well as of the
nuclear  motion part. The avoided crossing with the \B\ state leads to very complex shapes of the \D\  PEC
and of the SO and electronic angular momentum coupling curves with the $A$ and $C$ states. The corresponding
SOCs of the $B$ and $D$ states  with the nearby state $C$ are also relatively large, $\sim$ 30~\cm\ and
80~\cm, respectively (see Fig.~\ref{f:socs_new}), and therefore important. Besides, the $D$ PEC is rather
shallow with the equilibrium in the vicinity of the avoided crossing point, which also complicates the
solution. An accurate description of the $B$ and $D$ curves would require diabatic representations before
attempting any empirical refinement by fitting to the experiment.
In all cases our simulations, while not perfect, show striking agreement with the observed spectra.

\begin{table}
  \caption{Five spectroscopic windows (\cm) used to compare five vibronic bands of YO ($B$ and $D$) in Fig.~\protect\ref{f:T=77K:2017}.
  Experiment is by \citet{17ZhZhZh} while Theory is from this work.}
   \label{t:five:bands}
  \centering
  \begin{tabular}{ccccrrr}
    \hline
    \hline
    & Experiment &  Theory   &   Band \\
    \hline
  I & 20714.5 - 20753.5 &   20715 - 20754   & $B$(0,0) \\
 II & 23078.5 - 23117   &   23073 - 23112   & $D$(0,1) \\
III & 23837.5 - 23874.5 &   23769 - 23806   & $D$(1,1) \\
 IV & 23934.5 - 23973   &   23934.5 - 23973 & $D$(0,0) \\
  V & 24689   - 24730   &   24625.5 - 24666 & $D$(1,0) \\
    \hline
    \hline
\end{tabular}
\end{table}

\begin{figure}[H]
	\includegraphics[width=0.5\textwidth]{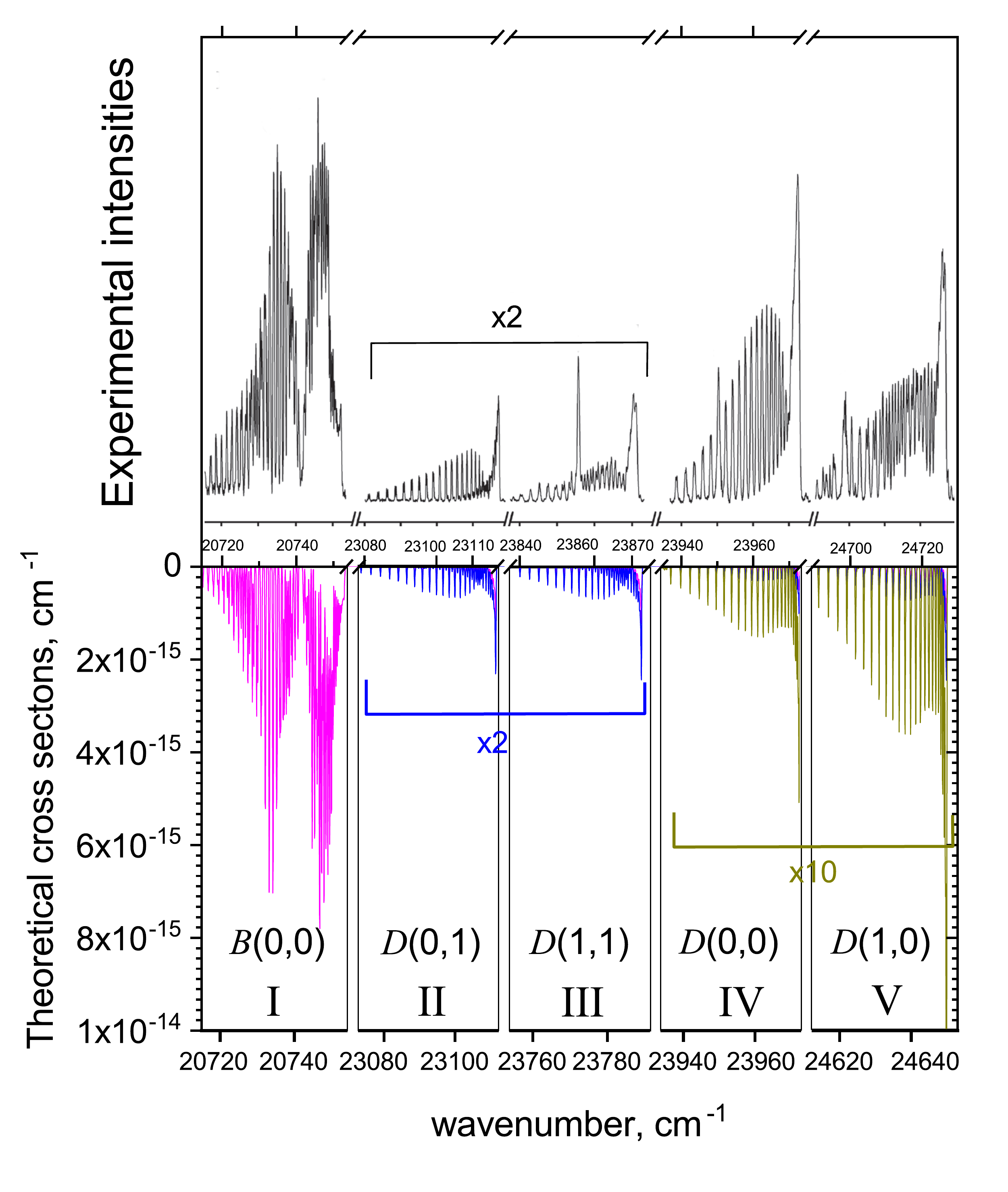}
	\caption{Comparison of our computed emission spectra to the measurements of \citet{17ZhZhZh}
	Our simulations assumed a cold rotational temperature of
	$T_{\rm rot}$ = 77 K and a hot vibrational temperature of $T_{\rm vib}$  = 2000~K. The Gaussian line profile of HWHM = 0.1~\cm\ was used. }
	\label{f:T=77K:2017}
\end{figure}

\subsection{Lifetimes}

The lifetimes of $^{89}$Y$^{16}$O in the \A\ and \B\ states ($v\le 2$) were measured by \citet{77LiPaxx.YO}
using laser fluorescence detection of nascent product state distributions in the reactions of Y with O$_2$,
NO, and SO$_2$. Some lifetimes were also measured by \citet{17ZhZhZh} and computed \ai\ by
\citet{88LaBaxx.YO}. Table~\ref{t:lf} presents a comparison of these results with our calculations with
\Duo,\cite{jt624} showing value of the states with corresponding lowest  $J$ and the positive parity. It can
be seen that our \A\ state lifetimes appear to be shorter than the observed ones. This suggests that the \A\
-- \X\  transition dipoles may be slightly too large. Good agreement is obtained for the lifetimes of the \B\
states, while the \D\ state lifetimes are underestimated by a factor of 2 indicating that the corresponding
transition dipole moments $D$--$B$ and $D$--$X$, or at least one of them might be too large.

\begin{table}
\caption{Lifetimes of $^{89}$Y$^{16}$O states in ns: comparison with the measurements of \citet{77LiPaxx.YO} and
\citet{17ZhZhZh}, and the \ai\ calculations of \citet{88LaBaxx.YO}. } \label{t:lf}
\begin{tabular}{lrcrrr}
\hline
\hline
State       &$v$ &   [\onlinecite{77LiPaxx.YO}]    & [\onlinecite{17ZhZhZh}]&  [\onlinecite{88LaBaxx.YO}] &   This work  \\
\hline
\A$_{1/2}$   &$         0  $&$     33.0 \pm 1. 3    $&$                    $&$                 21 $&$       22.7 $\\
             &$         1  $&$      36.5\pm 2.4     $&$                    $&$                    $&$       23.0 $\\
\A$_{3/2}$   &$         0  $&$     32. 3\pm 0. 9    $&$                    $&$                 21 $&$       20.9 $\\
             &$         1  $&$       30.4\pm1.8     $&$                    $&$                    $&$       21.3 $\\
             &$         2  $&$     33.4 \pm 1. 5    $&$                    $&$                    $&$       21.7 $\\
             &$         6  $&$      41.6\pm 2.1     $&$                    $&$                    $&$       35.7 $\\
\B\          &$         0  $&$      30.0\pm 0.9     $&$  38\pm 5           $&$                 17 $&$       26.7 $\\
             &$         1  $&$     32.5 \pm 1. 2    $&$                    $&$                    $&$       29.2 $\\
\D\          &$         0  $&$                      $&$      79 \pm 5      $&$                    $&$       34.1 $\\
\D\          &$         1  $&$                      $&$      79 \pm 5      $&$                    $&$       41.4 $\\
\hline
\end{tabular}
\end{table}

\section{Conclusion}

In this work, a composite approach to accurate first-principles description of the spectroscopy of open-shell
TM-containing diatomics is proposed and its high efficiency is demonstrated taking the example of the yttrium
oxide molecule. The approach is based on the combined use of single reference coupled cluster and
multireference methods of electronic structure theory, accompanied with a thorough joint analysis of the
SR/MR character of the molecular wave function. A full set of potential energy, (transition) dipole moment,
spin--orbit, and electronic angular momenta curves for the lowest 6 electronic states of YO was produced \ai\
using a combination of the CCSD(T)/CBS and MRCI methods. These curves were then used to solve the fully
coupled Schr\"{o}dinger equation for the nuclear motion using the \Duo\ program. Given the complexity of the
system under study, the results show remarkably good agreement with the experiment. Our ultimate goal is to
produce an accurate, empirical line list for $^{89}$Y$^{16}$O for applications in modelling the spectroscopy
of atmospheres of exoplanets and cool stars. This will require a refinement of the \ai\ curves by fitting to
the experimental data in the diabatic representation as well as inclusion of the non-adiabatic coupling
effects and will be addressed in future work. The \A\ band of YO has strong absorption in the visible region,
i.e. where the stellar radiation usually peaks. Such systems are known to cause the temperature inversion in
atmospheres of exoplanets, similar to the inversion caused by TiO and VO in giant exoplanets
\cite{17MaGaxx.TiO}. Opacities of such species are crucial in modelling the degree of temperature inversion
in giant exoplanets. YO is yet to be detected in exoplanetary atmospheres and this work is meant to provide
the necessary spectroscopic data.

YO is one of the few molecules with the strong potential for laser-cooling applications,\cite{18CoDiWu.YO}
which have widely ranging applications, from quantum information and chemistry to searches for new
fundamental physics. The results of this work will help to model the cooling properties of YO and thus will
be important for designing and implementing laser-cooling experiments.

The \ai\ curves of YO obtained in this study are provided as part of the supplementary material to this paper
along with our spectroscopic model in a form of a \Duo\ input file. The computed line list can be obtained
from \verb!www.exomol.com!. This is given in the ExoMol format \cite{jt631} which also includes
state-dependent lifetimes. The line list can be directly used with the {\sc ExoCross} program to simulate the
spectral properties of YO.

\section*{Acknowledgements}

This work was supported  by STFC Projects No. ST/M001334/1 and ST/R000476/1. The authors acknowledge the use
of the UCL Legion High Performance Computing Facility (Legion@UCL), and associated support services, in the
completion of this work, along with the Cambridge COSMOS SMP system, part of the STFC DiRAC HPC Facility
supported by BIS National E-infrastructure capital grant ST/P002307/1 and ST/R002452/1 and STFC operations
grant ST/R00689X/1. DiRAC is part of the National e-Infrastructure. A. N. S. and V. G. S. acknowledge support
from the Ministry of Science and Higher Education of the Russian Federation (Project No. 4.3232.2017/4.6).

\balance



\providecommand*{\mcitethebibliography}{\thebibliography}
\csname @ifundefined\endcsname{endmcitethebibliography}
{\let\endmcitethebibliography\endthebibliography}{}

\end{document}